\documentclass[11pt,prd,superscriptaddress,nofootinbib]{revtex4}
\usepackage{mathtext}
\usepackage{indentfirst}
\usepackage{epsfig,amsmath,amsfonts}

\usepackage{natbib}
\usepackage{graphicx}
\usepackage{verbatim}
\usepackage{amssymb}
\usepackage{amsmath}
\usepackage{amsfonts}
\usepackage{braket}
\usepackage[compat=1.1.0]{tikz-feynman}
\usepackage[english]{babel}
\usepackage{hyperref}
\hypersetup{urlcolor=red, colorlinks=true}  
\usepackage{pgfplots}
\usepackage{url}
\pgfplotsset{compat=1.12}
\usepackage{color}
\usepackage{vmargin}

\begin{document}

\hfill ITEP-TH-28/17

\hspace{2cm}

\centerline{\bf \LARGE A quantum heating as an alternative of reheating}

\vspace{10mm}

\centerline{{\bf Emil T. ${\rm \bf Akhmedov}^{1, 2}$} and {\bf Francesco ${\rm \bf Bascone}^{3}$}}

\vspace{5mm}

\begin{center}
{\it $\phantom{1}^{1}$ Institutskii per, 9, Moscow Institute of Physics and Technology, 141700, Dolgoprudny, Russia}
\end{center}

\begin{center}
{\it $\phantom{1}^{2}$ B. Cheremushkinskaya, 25, Institute for Theoretical and Experimental Physics, 117218, Moscow, Russia}
\end{center}

\begin{center}
{\it $\phantom{1}^{3}$ Dipartamento di Fisica e Chimica,
Università di Palermo, via Archirafi 36, 90123 Palermo - Italy}
\end{center}


\hspace{2cm}
\begin{center}
ABSTRACT
\end{center}

To model a realistic situation for the beginning we consider massive real scalar $\phi^4$ theory in a (1+1)-dimensional asymptotically static Minkowski spacetime with an intermediate stage of expansion. To have an analytic headway we assume that scalars have a big mass. At past and future infinities of the background we have flat Minkowski regions which are joint by the inflationary expansion region. We use the tree--level Keldysh propagator in the theory in question to calculate the expectation value of the stress--energy tensor which is, thus, due to the excitations of the zero--point fluctuations. Then we show that even for large mass, if the de Sitter expansion stage is long enough, the quantum loop corrections to the expectation value of the stress--energy tensor are not negligible in comparison with the tree--level contribution. That is revealed itself via the excitation of the higher--point fluctuations of the exact modes: During the expansion stage a non--zero particle number density for the exact modes is generated. This density is not Plankian and serves as a quench which leads to a thermalization in the out Minkowski stage.

\section{Introduction}

Standard reheating \cite{Kofman:1994rk} after inflation \cite{Starobinsky:1980te}, \cite{Guth:1980zm}, \cite{Linde:1981mu}, \cite{Albrecht:1982wi} demands the presence of inflaton field. One of the important ingredients of this picture is the belief that during the rapid expansion stage the matter in the unverse can only cool down. This is based on the common wisdom that quantum effects provide only very tiny contributions to such quantities as the stress--energy fluxes.

In fact, the so called one loop expectation value of the stress--energy tensor, which is calculated with the use of the tree--level two--point Hadamard function or Keldysh propagator usually does not provide strong contributions to the particle number density after inflation. The rapid expansion always wins. The non--trivial value of the stress--energy tensor, if any in such circumstances, is due to the excitation of the zero--point fluctuations, which are the only ones contributing to the tree--level two--point function in the ground state. At the same time, it is believed that quantum loop corrections to the expectation value under discussion provide only UV renormalization to physical quantities. However, this observation is based on the intuition gained in stationary vacuum quantum field theory loop calculations, which in generic situations are not applicable in such non--stationary situations as rapid inflationary expansion.

There is already a vast literature showing that quantum loop corrections during inflation can grow in time and become comparable with tree--level contributions. See e.g. \cite{MarolfMorrison}--\cite{Akhmedov:2013vka} for incomplete list of references. There are different sorts of secular effects. There are such effects which are specific only for the de Sitter space--time massless non--conformal scalars and gravitons (see e.g. \cite{Tsamis:2005hd}). The other secular effects are present in generic non--stationary situations and appear when the time separation between arguments of the two--point function is growing (see e.g. introduction of \cite{Akhmedov:2017ooy} for the review). The latter type of secular effects leads just to a mass renornalization or reveals a quasi--particle instability, if any. However, there are also such effects which may provide dramatic particle production \cite{PolyakovKrotov}--\cite{Akhmedov:2013vka} within \underline{comoving volume}. They appear in the two--point functions when both of their arguments are taken to the future infinity, while the time separation between them is held fixed. Of course the intensity of the particle production depends on their mass and on their initial density. However, even for very big mass and for vanishing initial density it may be comparatively relevant, if the expansion stage is long enough, as we show in the present note. It goes without saying that for light fields there can be even more dramatic effects \cite{Akhmedov:2017ooy}.

It is known in condensed matter theory that in non--stationary situations generally loop corrections can be strong \cite{LL}, \cite{Kamenev} for various reasons. Furthermore this fact has been observed in de Sitter space \cite{Akhmedov:2013vka}, \cite{Akhmedov:2017ooy} in strong electric field backgrounds \cite{Akhmedov:2014hfa}, \cite{Akhmedov:2014doa}, in the case of loop corrections to the Hawking radiation \cite{Akhmedov:2015xwa} and in the case of moving mirrors \cite{Akhmedov:2017hbj}, \cite{Alexeev:2017ath}. In all, even fundamental quantum fields in non--stationary situations reveal a similar behaviour to quantum fields in a medium, as in condensed matter theory.

The secular effect of interest for us in the present paper is due to the excitation of the higher levels (on top of the zero--point fluctuations) of the exact modes in background gravitational field. This is due to the non--stationarity of the inflationary expansion and due to the presence of interactions in the quantum field theory \cite{Akhmedov:2013vka} (see also \cite{Akhmedov:2017ooy}). In non--Gaussian (selfinteracting) theories in non--stationary situations initial ground state of the quantum field theory is changing in time that may reveal itself in dramatic changes of correlation functions, which are not observed in the proper vacuum quantum field theory measurements and calculations. In particular, the secular effects under discussion lead to a growth of the physical particle number density in time and may result in its non--zero value at the final stage of the expansion. That can be true even for very massive fields, if the expansion stage is long enough. As a result in this note we point out that heating can be achieved even without inflaton field, for any physical origin of the cosmological expansion and even for very massive fields.

For simplicity in this note we consider a model situation of the 2D $\phi^4$ quantum field theory in the gravitational background as follows. At past and future infinities of the background we have flat Minkowski regions which are joint by a long inflationary expansion stage.

The paper is organized as follows. In the section 2 we define the gravitational background and describe the properties of the free mode functions. In the section 3 we perform the standard calculation of the stress--energy expectation value with the use of the tree--level two--point Hadamard or Keldysh function. In the section 4 we calculate lowest loop corrections to the correlation functions and show that some of them grow with time. In the section 5 we perform the resummation of the leading corrections from all loops during the expansion stage and find the created particle density at the moment of exit from the expansion. In the section 6 we derive the kinetic equation in the final Minkowski stage and show that the created particle density serves as a quench for the thermalization. The section 7 contains conclusions.

\section{Setup of the problem}

The action of the theory that we consider in this paper is

\begin{equation}
S=\int d^2x \sqrt{|g|}\left(\frac{1}{2}g^{\mu\nu}\partial_\mu \phi\partial_\nu \phi-\frac{1}{2}m^2\phi^2-\frac{\lambda}{4!}\phi^4 \right),
\end{equation}
where the metric is that of (1+1)-dimensional Robertson-Walker spacetime:

\begin{equation}
ds^2=dt^2-a^2(t)dx^2 = C(\eta)(d\eta^2-dx^2),
\end{equation}
where $C(\eta)$ is the so called Conformal factor.

\begin{figure}[h!]
\centering
\includegraphics[scale=1]{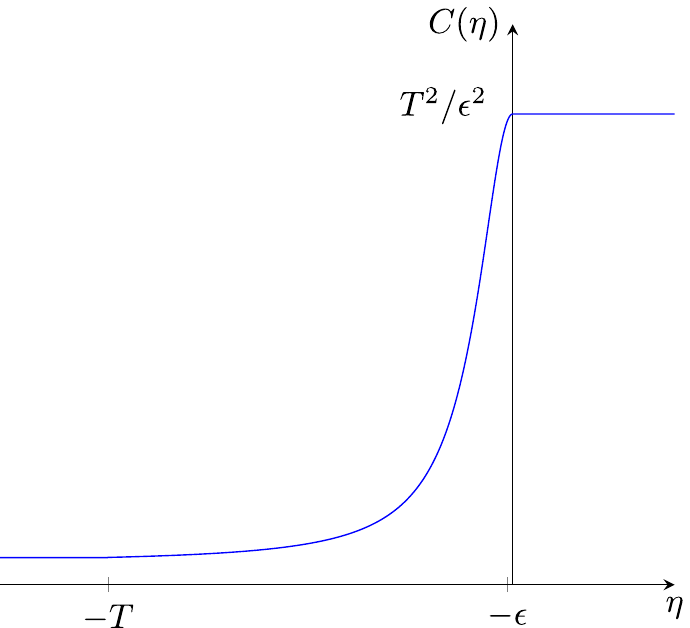}
\caption{Conformal factor for spacetime described by eq. (\ref{eq5})}
\label{fig:backgr}
\end{figure}

In particular we consider a (1+1)-dimensional asymptotically static universe with Minkowskian in-- and out--regions and with a de Sitter phase of expansion, with the following metric:

\begin{equation}
\label{eq5}
 ds^2 =
  \begin{cases}
   \left(1+\frac{T^2}{\eta^2+\epsilon^2} \right)\left[d\eta^2-dx^2 \right], & \text{} \eta \in (-\infty,0] \\
\frac{T^2}{\epsilon^2}\left[d\eta^2-dx^2 \right],   & \text{} \eta \in [0,+\infty),
  \end{cases}
  \quad {\rm where} \quad T^2 \ggg \epsilon^2
\end{equation}
This metric describes three phases (see fig. \ref{fig:backgr}):

\begin{enumerate}

\item First --- in-- Minkowski stage

\begin{equation}
ds^2\approx d\eta^2-dx^2 \approx dt^2-dx^2, \quad t \approx \eta \quad {\rm and} \quad \eta \ll -T,
\end{equation}

\item de Sitter expansion stage

\begin{equation}
ds^2 \approx \frac{T^2}{\eta^2}\left(d\eta^2-dx^2 \right), \quad - T < \eta < - \epsilon \quad {\rm and} \quad \epsilon \ll \left|\eta\right| \ll T,
\end{equation}

\item Second --- out-- Minkowski stage

\begin{equation}
ds^2 \approx \frac{T^2}{\epsilon^2}\left(d\eta^2-dx^2 \right) \approx dt^2-\frac{T^2}{\epsilon^2}dx^2,  \quad t \approx \frac{T}{\epsilon} \eta \quad {\rm and} \quad - \epsilon < \eta < 0, \quad \left|\eta \right| \ll \epsilon.
\end{equation}
We assume that there is the infinite flat region with the metric

\begin{equation*}
ds^2 \approx \frac{T^2}{\epsilon^2}\left(d\eta^2-dx^2 \right) \approx dt^2-\frac{T^2}{\epsilon^2}dx^2, \quad \eta \in [0,+\infty) \quad {\rm and} \quad t \in [0,+\infty),
\end{equation*}
\end{enumerate}
which is glued to the last stage.
Physically we can think that at some moment of time a cosmological constant was somehow created resulting in a de Sitter expansion that stops in a while, leaving the space flat again. The origin of this cosmological constant is irrelevant for the discussion in the present paper.

For the de Sitter expansion stage the scale factor has the exponential form $a(t) \approx e^{2Ht}$. In our case
the Hubble expansion rate is $H \approx \frac{1}{T}$.

\subsection{Modes in the expansion stage}

During the expansion stage the equations of motion for the free modes in the theory under consideration ($\lambda = 0$) are as follows:

\begin{equation}
\left(\eta^2\partial^2_\eta-\eta^2\partial^2_x+m^2T^2 \right)\phi(\eta,x) \approx 0.
\end{equation}
One can represent the modes which solve this equation as $\phi_k(\eta,x)=g_k(\eta)e^{\mp ikx}$ and the equation of motion for their temporal part, $g_k(\eta)$, is then

\begin{equation}
\frac{d^2g_k(\eta)}{d\eta^2}+\left(k^2+\frac{m^2T^2}{\eta^2} \right)g_k(\eta) \approx 0.
\end{equation}
If one assumes the ansatz $g_k(\eta)=|\eta|^{1/2}h(k\eta)$, then we obtain the Bessel equation for $h(k\eta)$ with the index $\nu=i\mu=i\sqrt{m^2T^2-\frac{1}{4}}$. Hence, one can write that \cite{Bunch:1978yq}:

\begin{equation}
h_k(\eta)=C_1 \, H^{(1)}_\nu(k|\eta|) + C_2 \, H^{(2)}_\nu(k|\eta|),
\end{equation}
where $H^{(1)}$ and $H^{(2)}$ are Hankel functions of the first and second kind and $C_1$,$C_2$ are some complex coefficients that we will fix by gluing solutions at $\eta=-T$ and by the normalization conditions.

Hankel functions with such an index behave as follows:

\[
 H_{\nu}^{(1),(2)}(x) \propto
  \begin{cases}
  \frac{e^{\pm ix}}{\sqrt{x}}, & \text{} x \gg |\nu| \\
A_+ x^{\nu} + A_- x^{-\nu},       & \text{} x \ll |\nu|,
  \end{cases}
\]
for some complex constants $A_\pm$.

If the field is heavy, $m>1/2T$, in de Sitter space it is said to belong to the principal series, while the light field, $m<1/2T$, belongs to the complementary series. Modes of the principal series oscillate and decay to zero as $\eta^{1/2\pm i\mu}$ when $\eta \rightarrow 0$, while for the light fields from the complementary series modes homogeneously decay to zero as $\eta^{1/2 \pm \sqrt{1/4-m^2T^2}}$, when $\eta \rightarrow 0$. In the present paper we work under the assumption of the heavy fields, $m \gg 1/2T$, so that $\nu \equiv i\mu$ is purely imaginary. It is understood in this case that we are talking about the analogue of principal and complementary series of de Sitter space as they are representations of the de Sitter isometry group that is not present in our case. So for this reason from now on we will just refer to heavy and light fields.

The quantum field can be expanded in the usual way:
\begin{equation}
\phi(\eta,x)=\int dk\, |\eta|^{1/2}\left[a_k e^{-ikx}h(k\eta)+a^\dagger_k e^{ikx}h^*(k\eta) \right].
\end{equation}
Annihilation, $a_k$, and creation, $a^{\dagger}_k$, operators obey the standard commutation relations as a corollary of the fact that $g_k(\eta)$ obeys the Klein-Gordon equation.

Because of the metric that we are considering here, the free Hamiltonian depends on time. Hence, the system under consideration is in a non--stationary state and one has to apply the Schwinger-Keldysh (aka in-in, aka Closed Time Path) diagrammatic technique. An introduction to it can be found in \citep{LL}, \cite{Kamenev}.
Every particle in this formalism is described by the matrix propagator:

\begin{equation*}
D \equiv \bordermatrix{~ &  &  \cr
                   & iD^K & D^R \cr
                   & D^A & 0 \cr}
\end{equation*}
for which the entries are the Keldysh $D^K$, Retarded $D^R$ and Advanced $D^A$ propagators:

\begin{eqnarray}
D^K(x_1,x_2)=\frac{1}{2}\langle \left\{\phi(x_1),\phi(x_2) \right\} \rangle, \nonumber \\
D^R(x_1,x_2)=\theta(\eta_1-\eta_2)\langle \left[\phi(x_1),\phi(x_2) \right] \rangle \quad {\rm and} \quad
D^A(x_1,x_2)=-\theta(\eta_2-\eta_1)\langle \left[\phi(x_1),\phi(x_2) \right] \rangle.
\end{eqnarray}
In this paper we consider only spatially homogeneous states and due to the spatial homogeneity of the background it is convenient to perform the Fourier transformation of all quantities along the spatial direction:

\begin{equation*}
D^{K,R,A}(k|\eta_1,\eta_2)\equiv \int dx\, e^{ikx} D^{K,R,A}(\eta_1,x;\eta_2,0),
\end{equation*}
hence,

\begin{equation}\label{eq:19}
D^K(k|\eta_1,\eta_2) = (\eta_1\eta_2)^{1/2} \, \left\{h_k(\eta_1)h^*_k(\eta_2)\left[\frac{1}{2}+n(k) \right]+h_k(\eta_1)h_k(\eta_2)\kappa(k)+c.c.\right\},
\end{equation}
\begin{equation}
D^A_R(k|\eta_1,\eta_2) = \mp 2 \, \theta\left[\pm(\eta_1-\eta_2)\right] \, (\eta_1\eta_2)^{1/2} \, {\rm Im} \{h_k(\eta_1)h^*_k(\eta_2) \}.
\end{equation}
Here $n(k) \equiv \langle a^{\dagger}_ka_k \rangle$, $\kappa(k) \equiv \langle a_k a_{-k} \rangle$ and $\kappa^*(k) \equiv \langle a^+_k a^+_{-k} \rangle$ are the number density and anomalous quantum average for the exact modes with respect to the state under consideration.
In the free field theory $n(k)$ and $\kappa(k)$ do not change in time and, in particular, are vanishing if one starts in the vacuum state, $a_k \, | in\rangle = 0$, for all $k$. However, if one turns on interactions, then generally these quantities reappear and depend on time \cite{Akhmedov:2013vka}. Furthermore, if the anomalous quantum average vanishes, then $n(k) \, g_k \, g^*_k$ acquires the meaning of the particle number density per \underline{comoving volume}, given that in this case the free Hamiltonian of the theory is diagonal. This will be important for the physical interpretation of the kinetic equation approach that we will present later.

\subsection{Gluing of the modes}

The approximate behaviour of the modes in the entire spacetime under consideration is as follows:

\begin{equation}\label{eq:17}
 g^{in}_k(\eta) \approx
  \begin{cases}
  \frac{1}{\sqrt{\omega_{in}}}e^{i\omega_{in}\eta}, & \text{} \eta \ll -T \\
|\eta|^{1/2}\Bigl[A_k H^{(1)}_\nu(k|\eta|)+B_k H^{(2)}_\nu(k|\eta|) \Bigr],       & \text{} -T \ll \eta \ll -\epsilon \\
\frac{1}{\sqrt{\omega_{out}}}\left(C_k e^{i\omega_{out}\eta}+D_k e^{-i\omega_{out}\eta} \right), 		& \text{} \eta \gg -\epsilon
  \end{cases}
\end{equation}
where $\omega_{in}(k)=\sqrt{k^2+m^2}$ and $\omega_{out}(k)=\sqrt{k^2+m^2\frac{T^2}{\epsilon^2}}$. This can be seen from the approximate form of the Klein--Gordon equation in the corresponding parts of the space--time.

To estimate the coefficients $A,B,C,D$ we need to glue the solutions and their first derivatives at $\eta=-T$ and $\eta=-\epsilon$. We start with the gluing at $\eta = -T$:

\begin{equation}
  \begin{cases}
  g^{M}_k(-T) \approx g^{dS}_k(-T), & \text{}  \\
\partial_\eta g^M_k(\eta)|_{\eta=-T} \approx \partial_\eta g^{dS}_k(\eta)|_{\eta=-T}.     & \text{}
  \end{cases}
\end{equation}
For the large momenta, when $k|\eta| \gg \mu$ for all $\eta \in [-T,-\epsilon]$ (note that in this case both $kT\gg\mu$ and $k\epsilon \gg \mu$) we have that:

\begin{equation}
  \begin{cases}
  \sqrt{\frac{k}{\omega_{in}(k)}}e^{-i\omega_{in}(k)T} \approx Ae^{ikT}+Be^{-ikT}, & \text{}  \\
-\sqrt{\frac{\omega_{in}(k)}{k}}e^{-i\omega_{in}(k)T} \approx Ae^{ikT}-Be^{-ikT},    & \text{}
  \end{cases}
\end{equation}
and from this system of equations we find the coefficients:

\begin{eqnarray}
A \equiv A^b_k \approx \frac{e^{-i(\omega_{in}+k)T}}{2}\left(\sqrt{\frac{k}{\omega_{in}}}-\sqrt{\frac{\omega_{in}}{k}} \right) \quad {\rm and} \quad
B \equiv B^b_k \approx \frac{e^{-i(\omega_{in}-k)T}}{2}\left(\sqrt{\frac{k}{\omega_{in}}}+\sqrt{\frac{\omega_{in}}{k}} \right),
\end{eqnarray}
where the index $b$ indicates that the coefficients are valid in the large momenta approximation.

Since we are interested in the very heavy case $mT \gg 1$, for high momentum $kT \gg \mu=\sqrt{\left(mT\right)^2 - \frac14} \approx mT$ so that $k \gg m$ and hence $\omega_{in}(k)\approx k$. So we have

\begin{equation}
\label{eq:23}
A^b_k \approx 0 \quad {\rm and} \quad B^b_k \approx 1.
\end{equation}
For low momenta, when $k|\eta| \ll \mu$ for all $\eta \in [-T,-\epsilon]$ (note that in this case both $kT\ll\mu$ and $kT \ll \mu$), we have at $\eta \approx - T$ that:

\[
  \begin{cases}
  \sqrt{\frac{1}{\omega_{in}(k)T}}e^{-i\omega_{in}(k)T} \approx A(kT)^{i\mu}+B(kT)^{-i\mu}, & \text{}  \\
-i\sqrt{\omega_{in}(k)}e^{-i\omega_{in}(k)T} \approx \frac{1}{2}T^{-1/2} \left[A(kT)^{i\mu}+B(kT)^{-i\mu} \right]+T^{1/2} \left[iAk\mu(kT)^{i\mu-1}-iBk\mu(kT)^{-i\mu} \right].    & \text{}
  \end{cases}
\]
Solving this system of equations, we find the coefficients

\begin{equation*}
A^s_k \approx \frac{-i\sqrt{\omega_{in}T}e^{-i\omega_{in}T}(kT)^{-i\mu}-\frac{1}{\sqrt{\omega_{in}}}e^{-i\omega_{in}T}\left[\frac{1}{2\sqrt{T}}(kT)^{-i\mu}-i\sqrt{T}k\mu(kT)^{-i\mu-1} \right]}{2i\mu},
\end{equation*}
\begin{equation}
B^s_k \approx \frac{i\sqrt{\omega_{in}T}e^{-i\omega_{in}T}(kT)^{i\mu}-\frac{1}{\sqrt{\omega_{in}}}e^{-i\omega_{in}T}\left[\frac{1}{2\sqrt{T}}(kT)^{i\mu}+i\sqrt{T}k\mu(kT)^{i\mu-1} \right]}{2i\mu}.
\end{equation}
For later convenience, let us write the coefficients as follows:

\begin{equation}
A^s_k=k^{-\nu}\alpha^s_k \quad B^s_k=k^{\nu}\beta^s_k,
\end{equation}
where
\begin{equation*}
\alpha^s_k \approx \frac{e^{-i\omega_{in}T}}{2\mu}\left[-\sqrt{\omega_{in}}T^{-i\mu+1/2}+\frac{1}{\sqrt{\omega_{in}}}\mu T^{-i\mu-1/2}+\frac{i}{2\sqrt{\omega_{in}}}T^{-i\mu-1/2} \right],
\end{equation*}
\begin{equation}
\beta^s_k \approx \frac{e^{-i\omega_{in}T}}{2\mu}\left[\sqrt{\omega_{in}}T^{i\mu+1/2}-\sqrt{\frac{1}{\omega_{in}}}\mu T^{i\mu-1/2}+\frac{i}{2\sqrt{\omega_{in}}}T^{i\mu-1/2} \right].
\end{equation}
Since we are working in the approximation that $mT \gg 1$ for small momentum $kT\ll\mu=\sqrt{m^2T^2-\frac{1}{4}}\approx mT$ and this leads to $k \ll m$ so that in the leading approximation $\omega_{in}(k)\approx m$ and

\begin{equation}
\alpha^s_k \approx i\frac{e^{-imT}}{4(mT)^{3/2}}T^{-i\mu}, \quad \beta^s_k \approx i\frac{e^{-imT}}{4(mT)^{3/2}}T^{i\mu}.
\end{equation}
For intermediate momenta, when $k\epsilon \ll \mu \ll kT$, we can distinguish between two regions: $k|\eta| > \mu$ and $k|\eta| < \mu$.
Hence, at $\eta=-\epsilon$, the modes behave as $H_{i\mu} \propto (k|\eta|)^{\pm i\mu}$, while at $\eta=-T$ we have that $H_{i\mu} \propto \frac{e^{\pm ik|\eta|}}{\sqrt{k|\eta|}}$. For the intermediate momenta in the gluing at $\eta=-T$ we find that
\begin{equation}
A^{int}_k \approx A^b_k \quad {\rm and} \quad B^{int}_k \approx B^b_k.
\end{equation}
We continue with the gluing of the modes at $\eta = - \epsilon$:

\begin{equation}
\begin{cases}
g^{M}_k(-\epsilon) \approx g^{dS}_k(-\epsilon) & \text{}  \\
\partial_\eta g^M_k(\eta)|_{\eta=-\epsilon} \approx \partial_\eta g^{dS}_k(\eta)|_{\eta=-\epsilon}.     & \text{}
\end{cases}
\end{equation}
We can then calculate the coefficients $C_k$ and $D_k$ from eq. (\ref{eq:17}) in complete analogy to the previous case.
On this side, for large momenta, $k|\eta| \gg \mu $ for all $ \eta \in [-T,-\epsilon]$, we have that:

\begin{equation}
  \begin{cases}
  C^b_ke^{i\omega_{out}\epsilon}+D^b_ke^{-i\omega_{out}\epsilon} \approx\sqrt{\frac{\omega_{out}}{k}}\left[A^b_ke^{ik\epsilon}+B^b_ke^{-ik\epsilon} \right], & \text{}  \\
C^b_ke^{i\omega_{out}\epsilon}-D^b_ke^{-i\omega_{out}\epsilon} \approx \sqrt{\frac{\omega_{out}}{k}}\left[A^b_ke^{ik\epsilon}-B^b_ke^{-ik\epsilon} \right]. & \text{}
  \end{cases}
\end{equation}
From this system of equations we find the coefficients:

\begin{equation*}
C^b_k \approx \sqrt{\frac{\omega_{out}}{k}}A^b_ke^{i\epsilon(k-\omega_{out})} = \sqrt{\frac{\omega_{out}}{k}}\left[\frac{e^{-i(\omega_{in}+k)T}}{2}\left(\sqrt{\frac{k}{\omega_{in}}} - \sqrt{\frac{\omega_{in}}{k}} \right) \right]e^{i\epsilon(k-\omega_{out})},
\end{equation*}
\begin{equation}
D^b_k \approx \sqrt{\frac{\omega_{out}}{k}}B^b_ke^{-i\epsilon(k-\omega_{out})}=\sqrt{\frac{\omega_{out}}{k}}\left[\frac{e^{-i(\omega_{in}-k)T}}{2}\left(\sqrt{\frac{k}{\omega_{in}}}+\sqrt{\frac{\omega_{in}}{k}} \right) \right] e^{-i\epsilon(k-\omega_{out})}.
\end{equation}\\
When $mT \gg 1$ and $k\epsilon, kT \gg \mu$ we can use the approximations $k\epsilon \gg \mu \approx mT$, hence, $k\gg \frac{mT}{\epsilon} $ so that $\omega_{out}(k)=\sqrt{k^2+\left(\frac{mT}{\epsilon}\right)} \approx k$ as well as $\omega_{in}(k) \approx k$.
In this case we can approximate the coefficients $C^b_k$ and $D^b_k$ as $C^b_k \approxeq 0 \quad {\rm and} \quad D^b_k \approxeq 1$.

For low momenta, $k|\eta|\ll \mu $ for all $\eta \in [-T,-\epsilon]$, we have that:

\begin{equation}
  \begin{cases}
  C^s_ke^{i\omega_{\text{out}}\epsilon}+D^s_ke^{-i\omega_{\text{out}}\epsilon} \approx \sqrt{\omega_{\text{out}}\epsilon}\left[A^s_k (k\epsilon)^{i\mu}+B^s_k(k\epsilon)^{-i\mu}\right] & \text{}  \\
 \begin{split}  C^s_ke^{i\omega_{\text{out}}\epsilon}-D^s_ke^{-i\omega_{\text{out}}\epsilon} & {} \approx  A^s_k\left[-\frac{i}{2\sqrt{\omega_{\text{out}}\epsilon}}(k\epsilon)^{i\mu}+k\mu\sqrt{\frac{\epsilon}{\omega_{\text{out}}}}(k\epsilon)^{i\mu-1} \right]+ \\ &{} -B^s_k \left[\frac{i}{2\sqrt{\omega_{\text{out}}\epsilon}}(k\epsilon)^{-i\mu}+k\mu\sqrt{\frac{\epsilon}{\omega_{\text{out}}}}(k\epsilon)^{-i\mu-1} \right]. \end{split}
  \end{cases}
\end{equation}
Solving the system we find the coefficients

\begin{equation*}
\begin{aligned}
C^s_k \approx {} & \frac{e^{-i\omega_{out}\epsilon}}{2}A^s_k\left[(k\epsilon)^{i\mu}\left(\sqrt{\omega_{out}\epsilon}-\frac{i}{2\sqrt{\omega_{out}\epsilon}} \right)+k\mu\sqrt{\frac{\epsilon}{\omega_{out}}}(k\epsilon)^{i\mu-1} \right]+ \\ & +\frac{e^{-i\omega_{out}\epsilon}}{2}B^s_k \left[(k\epsilon)^{-i\mu}\left(\sqrt{\omega_{out}\epsilon}-\frac{i}{2\sqrt{\omega_{out}\epsilon}} \right)-k\mu\sqrt{\frac{\epsilon}{\omega_{out}}}(k\epsilon)^{-i\mu-1} \right],
\end{aligned}
\end{equation*}
\begin{equation}
\begin{aligned}
D^s_k \approx {} & \frac{e^{i\omega_{out}\epsilon}}{2}A^s_k\left[(k\epsilon)^{i\mu}\left(\sqrt{\omega_{out}\epsilon}+\frac{i}{2\sqrt{\omega_{out}\epsilon}} \right)-k\mu\sqrt{\frac{\epsilon}{\omega_{out}}}(k\epsilon)^{i\mu-1} \right]+ \\ & +\frac{e^{i\omega_{out}\epsilon}}{2}B^s_k \left[(k\epsilon)^{-i\mu}\left(\sqrt{\omega_{out}\epsilon}+\frac{i}{2\sqrt{\omega_{out}\epsilon}} \right)+k\mu\sqrt{\frac{\epsilon}{\omega_{out}}}(k\epsilon)^{-i\mu-1} \right].
\end{aligned}
\end{equation}\\
When $mT \gg 1$ and $kT, k\epsilon \ll \mu$ we can use the approximations $k\epsilon \ll \mu \approx mT$, hence, $k\ll \frac{mT}{\epsilon}$ so that $\omega_{out}(k)=\sqrt{k^2+\left(\frac{mT}{\epsilon}\right)} \approxeq \frac{mT}{\epsilon}$ as well as $\omega_{in}(k) \approx m$.
In this case we can approximate the coefficients $C^s_k$ and $D^s_k$ as

\begin{equation*}
C^s_k \approx i\frac{e^{-2imT}}{8(mT)^{3/2}}\left[\left(\frac{T}{\epsilon} \right)^{-imT} \left(2\sqrt{mT}-\frac{i}{2\sqrt{mT}} \right)-\frac{i}{2\sqrt{mT}}\left(\frac{T}{\epsilon} \right)^{imT} \right] \approx i\frac{e^{-2imT}}{4(mT)}\left(\frac{T}{\epsilon} \right)^{-imT}
\end{equation*}
\begin{equation}
D^s_k \approx \frac{i}{8(mT)^{3/2}}\left[\left(\frac{T}{\epsilon} \right)^{imT}\left(2\sqrt{mT}+\frac{i}{2\sqrt{mT}} \right)+\frac{i}{\sqrt{mT}}\left(\frac{T}{\epsilon} \right)^{-imT} \right]\approx i\frac{1}{4(mT)}\left(\frac{T}{\epsilon} \right)^{imT}.
\end{equation}\\
Finally, in the intermediate region of momenta, for the $\eta=-\epsilon$ gluing, we find the following relations:

\begin{equation*}
\begin{aligned}
C^{int}_k \approx {} & \frac{e^{-i\omega_{out}\epsilon}}{2}A^b_k\left[(k\epsilon)^{i\mu}\left(\sqrt{\omega_{out}\epsilon}-\frac{i}{2\sqrt{\omega_{out}\epsilon}} \right)+k\mu\sqrt{\frac{\epsilon}{\omega_{out}}}(k\epsilon)^{i\mu-1} \right]+ \\ & -\frac{e^{-i\omega_{out}\epsilon}}{2}B^b_k \left[(k\epsilon)^{-i\mu}\left(\sqrt{\omega_{out}\epsilon}+\frac{i}{2\sqrt{\omega_{out}\epsilon}} \right)+k\mu\sqrt{\frac{\epsilon}{\omega_{out}}}(k\epsilon)^{-i\mu-1} \right],
\end{aligned}
\end{equation*}
\begin{equation}
\begin{aligned}
D^{int}_k \approx {} & \frac{e^{i\omega_{out}\epsilon}}{2}A^b_k\left[(k\epsilon)^{i\mu}\left(\sqrt{\omega_{out}\epsilon}+\frac{i}{2\sqrt{\omega_{out}\epsilon}} \right)-k\mu\sqrt{\frac{\epsilon}{\omega_{out}}}(k\epsilon)^{i\mu-1} \right]+ \\ & +\frac{e^{i\omega_{out}\epsilon}}{2}B^b_k \left[(k\epsilon)^{-i\mu}\left(\sqrt{\omega_{out}\epsilon}+\frac{i}{2\sqrt{\omega_{out}\epsilon}} \right)+k\mu\sqrt{\frac{\epsilon}{\omega_{out}}}(k\epsilon)^{-i\mu-1} \right],
\end{aligned}
\end{equation}
which in the limit we are working with, may be further approximated as follows:
\begin{equation}
C^{int}_k \approx -e^{-imT}(k\epsilon)^{-imT}\sqrt{mT}, \quad {\rm and} \quad
D^{int}_k \approx e^{-imT}(k\epsilon)^{-imT}\sqrt{mT}.
\end{equation}
We will need to perform the gluing of the modes with the use of the de Sitter out Jost modes in the section 6 to look at the out Minkowski region. The reason for that will be explained below.

\section{Tree--level expectation value of the stress--energy tensor}

We now evaluate the tree--level flux as this is usually associated with particle creation in such a situation as we discuss here.
The expectation values of the stress-energy tensor components can be calculated with the use of the well known relation

\begin{equation}
\bra{0}T_{\mu\nu}\ket{0}=\int dk \, T_{\mu\nu}[u_k,u^*_k],
\end{equation}
where $T_{\mu\nu}[u_k,u^*_k]$ denotes the bilinear expression for the energy momentum tensor in terms of the mode functions $u_k$ \cite{BirDav}.
To calculate the particle flux we need to use the in-modes in the out region:

\begin{equation}
u^{in}_k \approx \frac{1}{\sqrt{4\pi\omega_{out}(k)}}e^{ikx}\left(C_k e^{i\omega_{out}(k)\eta}+D_ke^{-i\omega_{out}(k)\eta} \right),
\end{equation}
because we are interested in the value of $\bra{in}T_{\mu\nu}\ket{in}$ as $t \to + \infty$, where $|in\rangle$ is the ground state with respect to the in--modes. The latter we consider as the initial state of the problem under consideration.

We can then calculate the integral defining $\bra{in}T_{\mu\nu}\ket{in}$ with the use of the asymptotic expressions for the coefficients $C_k$ and $D_k$ that we have found in the previous section. To do that we separate the integration into three regions of low, intermediate and high momenta:

\begin{equation}
\int dk \, T_{\mu\nu}[u_k,u^*_k]=\sum_{i=1}^{3} \int_{\Omega_i} dk \, T_{\mu\nu}[u_k,u^*_k],
\end{equation}
where the regions of integration, $\Omega_i$, are defined as follows

\begin{equation*}
\Omega_1=\left\{k : |k\eta| \ll \mu, \quad {\rm for \,\, all} \quad \eta \in [-T,-\epsilon] \right\},
\end{equation*}
\begin{equation*}
\Omega_2=\left\{k : |k|\epsilon<\mu<|k|T \right\},
\end{equation*}
\begin{equation*}
\Omega_3=\left\{k : |k\eta| \gg \mu, \quad {\rm for \,\, all} \quad \eta \in [-T,-\epsilon] \right\}.
\end{equation*}
The flux is then given by the expectation value of the non-diagonal components of the stress-energy tensor

\begin{equation}
\begin{aligned}
\bra{in}T_{\eta x}\ket{in} =\bra{in}T_{x\eta}\ket{in} = \int dk \partial_\eta u_k \partial_x u^*_k \\ \approx \int dk\left[\frac{i\omega_{out}}{\sqrt{4\pi\omega_{out}}}e^{ikx}\left(D_k e^{i\omega_{out}\eta}-C_k e^{-i\omega_{out}\eta}\right)\frac{-ik}{\sqrt{4\pi\omega_{out}}}e^{-ikx}\left(C^*_k e^{i\omega_{out}\eta}+D^*_k e^{-i\omega_{out}\eta} \right) \right] \\ \approx \int dk\frac{k}{4\pi}\left(C^*_k D_k e^{2i\omega_{out}\eta}+|D_k|^2-|C_k|^2-C_k D^*_k e^{-2i\omega_{out}\eta}  \right) \\ \approx \frac{1}{4\pi}\Bigl[\left(\frac{e^{2imT\left(1+\frac{\eta}{\epsilon}\right)}}{(4mT)^2}\left(\frac{T}{\epsilon}\right)^{2imT} - \frac{e^{-2imT\left(1+\frac{\eta}{\epsilon}\right)}}{(4mT)^2}\left(\frac{T}{\epsilon}\right)^{-2imT}+1 \right)\int_{\Omega_1}dk \, k + \\ + mT\sin\left(\frac{2mT}{\epsilon} \right)\int_{\Omega_2}dk \, k +\int_{\Omega_3}dk \, k \Bigr] = 0.
\end{aligned}
\end{equation}
The last equality follows from the fact that the integrands are odd functions in symmetric integration intervals. In principle one could obtain non--zero fluxes from the momenta regions $\Omega_1$ and $\Omega_2$, if the situation under consideration would not have been spatially homogeneous. In spatially homogeneous situations there are fluxes in both directions, which compensate each other. From the region $\Omega_3$ we obtain the standard expression as in flat space, because UV modes are not sensitive to the background curvature.

For the other components of the stress-energy tensor we find:

\begin{equation*}
\begin{aligned}
\bra{in}T_{\eta\eta}\ket{in} \approx {} & \frac{1}{2}\int dk \left[ \partial_\eta u\partial_\eta u^*+\partial_x u\partial_x u^*+m^2C(\eta \to + \infty) \, uu^* \right] \\ & \approx \frac{1}{2}\int dk \frac{\omega_{out}}{4\pi} \left[|C_k|^2+|D_k|^2-\left(C_k D^*_k e^{-2i\omega_{out}\eta}+C^*_k D_k e^{2i\omega_{out}\eta}\right)\right] \\ & +\frac{1}{2}\int dk\frac{k^2}{4\pi\omega_{out}}\left[|C_k|^2+|D_k|^2+\left(C_k D^*_k e^{-2i\omega_{out}\eta}+C^*_k D_k e^{2i\omega_{out}\eta}\right) \right] \\ & +\frac{1}{2}\int dk\frac{\left(\frac{mT}{\epsilon}\right)^2}{4\pi\omega_{out}}\left[|C_k|^2+|D_k|^2+\left(C_k D^*_k e^{-2i\omega_{out}\eta}+C^*_k D_k e^{2i\omega_{out}\eta}\right) \right] \\ & \approx \frac{1}{8\pi}\int dk \left(|C_k|^2+|D_k|^2 \right)\left(\omega_{out}+\frac{k^2}{\omega_{out}}+\frac{\left(\frac{mT}{\epsilon}\right)^2}{\omega_{out}}\right)\\ & + \frac{1}{8\pi}\int dk \left(C_k D^*_k e^{-2i\omega_{out}\eta}+C^*_k D_k e^{2i\omega_{out}\eta}\right)\left(-\omega_{out}+\frac{k^2}{\omega_{out}}+\frac{\left(\frac{mT}{\epsilon}\right)^2}{\omega_{out}} \right) \\ & \approx \frac{1}{4\pi}\left(\frac{1}{8mT\epsilon}\int_{\Omega_1}dk + \frac{2(mT)^2}{\epsilon}\int_{\Omega_2}dk + \int_{\Omega_3} dk \, \left|k\right| \right),
\end{aligned}
\end{equation*}
and

\begin{equation*}
\begin{aligned}
\bra{in}T_{xx}\ket{in} \approx {} & \frac{1}{2}\int dk \left[ \partial_\eta u\partial_\eta u^*+\partial_x u\partial_x u^*-m^2C(\eta \to + \infty) uu^* \right] \\ & \approx \frac{1}{2}\int dk \frac{\omega_{out}}{4\pi} \left[|C_k|^2+|D_k|^2-\left(C_k D^*_k e^{-2i\omega_{out}\eta}+C^*_k D_k e^{2i\omega_{out}\eta}\right)\right] \\ & +\frac{1}{2}\int dk\frac{k^2}{4\pi\omega_{out}}\left[|C_k|^2+|D_k|^2+\left(C_k D^*_k e^{-2i\omega_{out}\eta}+C^*_k D_k e^{2i\omega_{out}\eta}\right) \right] \\ & -\frac{1}{2}\int dk\frac{\left(\frac{mT}{\epsilon}\right)^2}{4\pi\omega_{out}}\left[|C_k|^2+|D_k|^2+\left(C_k D^*_k e^{-2i\omega_{out}\eta}+C^*_k D_k e^{2i\omega_{out}\eta}\right) \right] \\ & \approx \frac{1}{4\pi}\left(\frac{\epsilon}{8(mT)^3}\int_{\Omega_1}dk \, k^2 + 2\epsilon\int_{\Omega_2}dk\, k^2 + \int_{\Omega_3} dk \, \left|k\right|\right).
\end{aligned}
\end{equation*}
In the last two expressions we obtain the standard UV divergences as in flat space--time, which are coming from the $\Omega_3$ region, as it should be.
By the normal ordering we can cancel these contributions. The rest are the expectation values that we are looking for.
In the next section we will calculate loop contributions to the two--point correlation functions, which will substantially correct the expressions found in this section.

\section{Two-loop contributions in de Sitter expansion stage}

In this section we discuss the leading infrared loop contributions to the retarded, advanced, Keldysh propagators and vertexes in the limit as $\eta_{1,2} \to -\epsilon$ and $\frac{T}{\epsilon} \rightarrow +\infty$. See, for example, \citep{vanderMeulen:2007ah} for the Schwinger--Keldysh diagrammatic technique in $\phi^4$ scalar field theory in cosmology.

It is straightforward to see that one loop bubble diagram corrections to the two--point functions lead just to a mass renormalisation (perhaps time dependent). They generally do not grow as $T/\epsilon \to \infty$. Also it is straightforward to see that retarded and advanced propagators do not receive growing with $T/\epsilon$ corrections at two--loop sunset diagram order \cite{Akhmedov:2013vka} (see also \cite{Kamenev} for a related general discussion). Furthermore, it can be shown that for massive fields vertexes also do not receive growing corrections \cite{Akhmedov:2013vka}. (Note that this is not true for very light fields \cite{Akhmedov:2017ooy}.) Thus, we continue with the two--loop sunset diagrams for the Keldysh propagator.

Leading contributions to the Keldysh propagator at 2-loop order $D^K_2(k|\eta_1,\eta_2)$, in the limit $\frac{T}{\epsilon} \rightarrow +\infty$, are contained in the following expression \cite{Akhmedov:2013vka}:

\begin{equation}
D^K_2(k|\eta_1,\eta_2)\approx g_k(\eta_1)g^*_k(\eta_2)n_2(k)+g_k(\eta_1)g_k(\eta_2)\kappa_2(k)+c.c.
\end{equation}
with
\begin{eqnarray}
n_2(k) \propto \lambda^2\iint_T^\epsilon d\eta_3 \, d\eta_4 \, g_k(\eta_3)g^*_k(\eta_4)F(k|\eta_3,\eta_4), \nonumber \\ {\rm and} \quad
\kappa_2(k) \propto -\lambda^2\int_T^\epsilon \int_T^{\eta_3} d\eta_3 \, d\eta_4 \, g^*_k(\eta_3)g^*_k(\eta_4)F(k|\eta_3,\eta_4), \label{eq41}
\end{eqnarray}
where

\begin{equation}
\label{eq:31}
F(k|\eta_3,\eta_4)=\iint\frac{dq_1}{2\pi}\frac{dq_2}{2\pi}C(\eta_3)C(\eta_4) g_{q_1}(\eta_3)g^*_{q_1}(\eta_4)g_{q_2}(\eta_3)g^*_{q_2}(\eta_4)g_{|k-q_1-q_2|}(\eta_3)g^*_{|k-q_1-q_2|}(\eta_4),
\end{equation}
and $g_k(\eta)$ is defined in eq. (\ref{eq:17}), while $C(\eta)$ is in eq. (\ref{eq5}). The subscript 2 of $D^K$, $n$, $\kappa$ and $\kappa^*$ indicates that we are discussing here only the second loop corrections to these quantities.

The above expressions appear from the following 2-loop correction to the Keldysh propagator:

\begin{equation}
\label{eq:32}
\begin{aligned}
{} & D^K_2(k|\eta_1,\eta_2) \approx \frac{\lambda^2}{6}\iint\frac{dq_1}{2\pi}\frac{dq_2}{2\pi}\iint_{-T}^{-\epsilon} d\eta_3 d\eta_4\, C(\eta_3)C(\eta_4) \times \\ & \times \bigg[-\frac{1}{4}D^R_0(k|\eta_1,\eta_3)D^R_0(q_1|\eta_3,\eta_4)D^R_0(q_2|\eta_3,\eta_4)D^R_0(|k-q_1-q_2||\eta_3,\eta_4)D^K_0(k|\eta_4,\eta_2)+ \\& -\frac{1}{4}D^K_0(k|\eta_1,\eta_3)D^A_0(q_1|\eta_3,\eta_4)D^A_0(q_2|\eta_3,\eta_4)D^A_0(|k-q_1-q_2||\eta_3,\eta_4)D^A_0(k|\eta_4,\eta_2)+ \\ & -\frac{3}{4}D^R_0(k|\eta_1,\eta_3)D^K_0(q_1|\eta_3,\eta_4)D^R_0(q_2|\eta_3,\eta_4)D^R_0(|k-q_1-q_2||\eta_3,\eta_4)D^A_0(k|\eta_4,\eta_2)+ \\ & -\frac{3}{4}D^R_0(k|\eta_1,\eta_3)D^K_0(q_1|\eta_3,\eta_4)D^A_0(q_2|\eta_3,\eta_4)D^A_0(|k-q_1-q_2||\eta_3,\eta_4)D^A_0(k|\eta_4,\eta_2)+ \\ & +3D^R_0(k|\eta_1,\eta_3)D^K_0(q_1|\eta_3,\eta_4)D^K_0(q_2|\eta_3,\eta_4)D^R_0(|k-q_1-q_2||\eta_3,\eta_4)D^K_0(k|\eta_4,\eta_2)+ \\ & +3D^K_0(k|\eta_1,\eta_3)D^K_0(q_1|\eta_3,\eta_4)D^K_0(q_2|\eta_3,\eta_4)D^A_0(|k-q_1-q_2||\eta_3,\eta_4)D^A_0(k|\eta_4,\eta_2)+ \\ & +D^R_0(k|\eta_1,\eta_3)D^K_0(q_1|\eta_3,\eta_4)D^K_0(q_2|\eta_3,\eta_4)D^K_0(|k-q_1-q_2||\eta_3,\eta_4)D^A_0(k|\eta_4,\eta_2) \bigg],
\end{aligned}
\end{equation}
if we neglect the difference between $\eta_1$, $\eta_2$ and $\eta=\sqrt{\eta_1\eta_2} \to - \epsilon$ and set the initial time to be $-T$. Such a neglection can be done if we keep only the largest contribution to $n$, $\kappa$ and $\kappa^*$ as $\frac{T}{\epsilon} \rightarrow +\infty$ \cite{Akhmedov:2013vka}.

The Schwinger--Keldysh diagrams defining eq. (\ref{eq:32}) are as follows:
\\ \\ \\ \\
\begin{figure}[h!]
\centering
\includegraphics[scale=1]{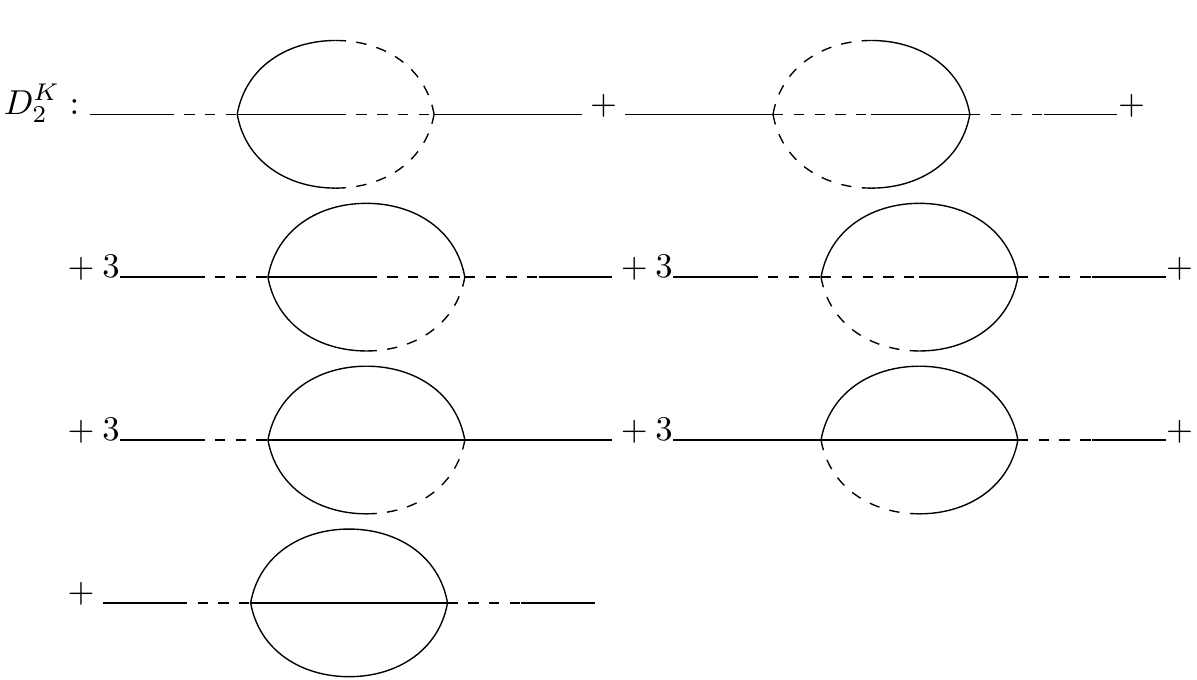}
\label{fig:diagrams}
\end{figure}

We start with 2-loop corrections to the Keldysh propagator for small external momenta, i.e. $k|\eta| \ll\mu$ for all $\eta \in [-T,-\epsilon]$. In this case

\begin{equation*}
\begin{aligned}
h_k(\eta_3)h^*_k(\eta_4)= {} & \left(\alpha^s_k \eta_3^{i\mu}+\beta^s_k \eta_3^{-i\mu} \right)\left(\alpha^{s*}_k \eta_4^{-i\mu}+\beta^{s*}_k \eta_4^{i\mu} \right) \\ & \approx |\alpha^s_k|^2\left(\frac{\eta_3}{\eta_4} \right)^{i\mu}+|\beta^s_k|^2\left(\frac{\eta_3}{\eta_4} \right)^{-i\mu} \propto \frac{(mT)^{-3}}{16} \left[\left(\frac{\eta_3}{\eta_4} \right)^{i\mu}+ \left(\frac{\eta_3}{\eta_4} \right)^{-i\mu}\right],
\end{aligned}
\end{equation*}
if subleading terms are neglected. Using this expression we can find from eq. (\ref{eq:31}) that $n_2$ is as follows

\begin{eqnarray}
n_2(k) \propto \frac{\lambda^2T^4}{(mT)^{3}}\iint^{-\epsilon}_{-T} d\eta_3 \, d\eta_4 \left[\left(\frac{\eta_3}{\eta_4}\right)^{i\mu}+\left(\frac{\eta_3}{\eta_4}\right)^{-i\mu} \right] \times \nonumber \\ \times \iint dq_1 dq_2 \, h_{q_1}(\eta_3)h^*_{q_1}(\eta_4)h_{q_2}(\eta_3)h^*_{q_2}(\eta_4)h_{q_3}(\eta_3)h^*_{q_3}(\eta_4),
\end{eqnarray}
where $q_3=|k-q_1-q_2|$. The largest contribution to this expression comes from the region where $q_{1,2,3} \gg k$ \cite{Akhmedov:2013vka}. (Note that this is true only for the massive fields \cite{Akhmedov:2017ooy}.)

To evaluate the last integral we make the following change of the integration variables:

\begin{equation*}
\eta_3 \rightarrow u=\sqrt{\eta_3\eta_4}, \quad \eta_4 \rightarrow v=\sqrt{\frac{\eta_3}{\eta_4}} \quad {\rm and} \quad q_i \rightarrow l_i=uq_i.
\end{equation*}
Then, it becomes equal to

\begin{eqnarray}
n_2(k) \propto \frac{\lambda^2T^4}{(mT)^{3}}\iint_T^\epsilon \iint \frac{du \, dv}{uv}dl_1 dl_2 \left(v^{2i\mu}+v^{-2i\mu} \right) \times \nonumber \\ \times h_{\frac{l_1}{u}}(uv)h^*_{\frac{l_1}{u}}\left(\frac{u}{v}\right)h_{\frac{l_2}{u}}(uv)h^*_{\frac{l_2}{u}}\left(\frac{u}{v}\right) h_{\frac{l_3}{u}}(uv)h^*_{\frac{l_3}{u}}\left(\frac{u}{v}\right).
\end{eqnarray}
The integration over the internal momenta, $q_{1,2}$, can be separated into four regions: The first one is when $|q_{1,2}| \ll \frac{\mu}{\sqrt{\eta_3\eta_4}}$, the second region is when $|q_{1,2}| \gg \frac{\mu}{\sqrt{\eta_3\eta_4}}$, the third region is when $|q_1| \ll \frac{\mu}{\sqrt{\eta_3\eta_4}}$ and $|q_2| \gg \frac{\mu}{\sqrt{\eta_3\eta_4}} $ and, finally, the fourth region is when $|q_1| \gg \frac{\mu}{\sqrt{\eta_3\eta_4}}$ and $|q_2| \ll \frac{\mu}{\sqrt{\eta_3\eta_4}}$. As a result,

\begin{equation*}
\begin{aligned}
n_2(k)  \propto \frac{\lambda^2T^4}{(mT)^{3}}{} & \Bigg[ \iint_T^\epsilon \iint^{\mu} \frac{du \, dv}{uv}dl_1 dl_2 \left(v^{2i\mu}+v^{-2i\mu} \right) \prod_{i=1}^3
h_{\frac{l_i}{u}}(uv)h^*_{\frac{l_i}{u}}\left(\frac{u}{v}\right)+ \\ & +\iint_T^\epsilon \iint_{\mu} \frac{du \, dv}{uv}dl_1 dl_2 \left(v^{2i\mu}+v^{-2i\mu} \right) \prod_{i=1}^3
h_{\frac{l_i}{u}}(uv)h^*_{\frac{l_i}{u}}\left(\frac{u}{v}\right)+ \\ & + \iint_T^\epsilon \int^{\mu}\int_{\mu} \frac{du \, dv}{uv}dl_1 dl_2 \left(v^{2i\mu}+v^{-2i\mu} \right) \prod_{i=1}^3
h_{\frac{l_i}{u}}(uv)h^*_{\frac{l_i}{u}}\left(\frac{u}{v}\right)+ \\ & + \iint_T^\epsilon \int_{\mu}\int^{\mu} \frac{du \, dv}{uv}dl_1 dl_2 \left(v^{2i\mu}+v^{-2i\mu} \right) \prod_{i=1}^3
h_{\frac{l_i}{u}}(uv)h^*_{\frac{l_i}{u}}\left(\frac{u}{v}\right) \Bigg].
\end{aligned}
\end{equation*}
For low internal momenta part of the integral, $|q_{1,2}|\ll \frac{\mu}{\sqrt{\eta_3\eta_4}}$, we have that:

\begin{equation*}
h_{q_i}(\eta_3)h^*_{q_i}(\eta_4) \propto (mT)^{-3} \left[\left(\frac{\eta_3}{\eta_4} \right)^{i\mu}+ \left(\frac{\eta_3}{\eta_4} \right)^{-i\mu}\right], \quad i=1,2,3,
\end{equation*}
and the corresponding contribution to $n_2$ is as follows:

\begin{eqnarray}
\frac{\lambda^2T^4}{(mT)^{3}}\iint_T^\epsilon \iint^{\mu} \frac{du \, dv}{uv}dl_1 dl_2 \left(v^{2i\mu}+v^{-2i\mu} \right) \prod_{i=1}^3
h_{\frac{l_i}{u}}(uv)h^*_{\frac{l_i}{u}}\left(\frac{u}{v}\right) \approx \nonumber \\ \approx \frac{\lambda^2T^4}{(mT)^{12}}\iint_T^\epsilon \iint^{\mu} \frac{du \, dv}{uv}dl_1 dl_2 \left(v^{2i\mu}+v^{-2i\mu} \right)^4= \nonumber \\
=\frac{\lambda^2}{m^4(mT)^{8}}\int_T^{\epsilon}\frac{du}{u}\int_T^{\epsilon}dv\iint^{\mu}dl_1dl_2f(v)=\frac{\lambda^2}{m^4}\log\left(\frac{\epsilon}{T} \right)\sigma_1(mT),
\end{eqnarray}
where $\sigma_1(mT)$ denotes the integral factor as a function of T and the logarithmic behaviour follows from the fact that the integrand of $du/u$ does not depend on $u$. This is the consequence of the fact that we have an approximate scaling symmetry, $\eta \to a \eta$ and $p \to p/a$, in the expansion stage, which is just a part of the de Sitter isometry group.

For large internal momenta part of the integral, $|q_{1,2}| \gg \frac{\mu}{\sqrt{\eta_3\eta_4}}$, we have that:

\begin{equation}
h_{q_i}(\eta_3)h^*_{q_i}(\eta_4)\propto \frac{e^{-iq_i\left(\eta_3-\eta_4 \right)}}{q_i\sqrt{\eta_3\eta_4}}, \quad i=1,2,3,
\end{equation}
because of the approximate form of the coefficient $A^b_q, B^b_q$ in eq. (\ref{eq:23}). The corresponding contribution to $n_2$ is as follows:

\begin{eqnarray}
\frac{\lambda^2T^4}{(mT)^{3}}\iint_T^\epsilon \iint_{\mu} \frac{du \, dv}{uv}dl_1 dl_2 \left(v^{2i\mu}+v^{-2i\mu} \right) \prod_{i=1}^3
h_{\frac{l_i}{u}}(uv)h^*_{\frac{l_i}{u}}\left(\frac{u}{v}\right) \approx \nonumber \\ \approx \frac{\lambda^2T^4}{(mT)^{3}}\iint_T^\epsilon \iint^{\mu} \frac{du \, dv}{uv}dl_1 dl_2 \left(v^{2i\mu}+v^{-2i\mu} \right)\frac{e^{-il_1\left(v-\frac{1}{v} \right)}}{l_1}\frac{e^{-il_2\left(v-\frac{1}{v} \right)}}{l_2}\frac{e^{-i|l_1+l_2|\left(v-\frac{1}{v} \right)}}{|l_1+l_2|}  \equiv \nonumber \\ \equiv \frac{\lambda^2 }{m^4}\log\left(\frac{\epsilon}{T} \right)\sigma_2(mT)
\end{eqnarray}
Hence, the region of high internal momenta also provides the logarithmic growth.

In the other two regions of integration ($|q_1| \ll \frac{\mu}{\sqrt{\eta_3\eta_4}}, |q_2| \gg \frac{\mu}{\sqrt{\eta_3\eta_4}} $ and $|q_1| \gg \frac{\mu}{\sqrt{\eta_3\eta_4}}, |q_2| \ll \frac{\mu}{\sqrt{\eta_3\eta_4}}$) we have a very similar situation, with a resulting growth of the form $\frac{\lambda^2}{m^4}\log\left(\frac{\epsilon}{T} \right)\sigma_{3,4}(mT)$, with some $\sigma_3 = \sigma_4$.

Thus, for low external momenta, $k|\eta| \ll \mu$, there is a logarithmic growth of $n_2(k)$ over the entire region of internal momenta:

\begin{equation}
n_2(k)  \approx \lambda^2\log \left(\frac{\epsilon}{T} \right)\sigma^{small}(mT),
\end{equation}
where $\sigma^{small}(mT)=\sum_{i=1}^{4}\sigma_i(mT)$ and in $\sigma_i(mT)$ we include T-dependent prefactors.

For the case of $\kappa_2(k)$ we have the same situation. The only difference is in the T-dependent prefactors. In this case we have that:

\begin{equation*}
\begin{aligned}
h^*_k(\eta_3)h^*_k(\eta_4)= {} & \left(\alpha^{s*}_k \eta_3^{-i\mu}+\beta^{s*}_k \eta_3^{i\mu} \right)\left(\alpha^{s*}_k \eta_4^{-i\mu}+\beta^{s*}_k \eta_4^{i\mu} \right) \\ & \approx \alpha^{s*}_k\beta^{s*}_k\left[\left(\frac{\eta_3}{\eta_4} \right)^{i\mu}+\left(\frac{\eta_3}{\eta_4} \right)^{-i\mu}\right] \propto -\frac{e^{2imT}}{16(mT)^3} \left[\left(\frac{\eta_3}{\eta_4} \right)^{i\mu}+ \left(\frac{\eta_3}{\eta_4} \right)^{-i\mu}\right].
\end{aligned}
\end{equation*}
As a result:

\begin{eqnarray}
\kappa_2(k) \propto \frac{\lambda^2T^2 e^{2imT}}{(mT)^{3}}\int_T^\epsilon du \int^1 dv \frac{1}{uv} \times \nonumber \\ \times \iint dl_1 dl_2 \left(v^{2i\mu}+v^{-2i\mu} \right) h_{\frac{l_1}{u}}(uv)h^*_{\frac{l_1}{u}}\left(\frac{u}{v}\right)h_{\frac{l_2}{u}}(uv)h^*_{\frac{l_2}{u}}\left(\frac{u}{v}\right) h_{\frac{l_3}{u}}(uv)h^*_{\frac{l_3}{u}}\left(\frac{u}{v}\right).
\end{eqnarray}
Hence, with the same manipulations as for $n_2$ we can just state that

\begin{equation}
\kappa_2(k)  \approx \lambda^2\log \left(\frac{\epsilon}{T} \right)\Gamma^{small}(mT),
\end{equation}
with a factor $\Gamma^{small}(mT)$ which follows from the previous equation.

Now let us look at the intermediate external momentum region, $k\epsilon\ll\mu$ and $kT \gg \mu$.
In this case we have that:

\begin{equation*}
h_k(\eta)= A^{int}_k H^{(1)}_{i\mu}+B^{int}_k H^{(2)}_{i\mu}\equiv A^{b}_k H^{(1)}_{i\mu}+B^{b}_k H^{(2)}_{i\mu} \approx C_+(k|\eta|)^{i\mu}+C_-(k|\eta|)^{-i\mu}.
\end{equation*}
Hence,

\begin{equation*}
h_k(\eta_3)h^*_k(\eta_4)\propto |C_+|^2\left(\frac{\eta_3}{\eta_4} \right)^{i\mu}+|C_-|^2\left(\frac{\eta_3}{\eta_4} \right)^{-i\mu},
\end{equation*}
and as a result,

\begin{equation*}
n_2(k) \propto \lambda^2T^2\iint_{\frac{\mu}{k}}^\epsilon \iint \frac{du \, dv}{uv}dl_1 dl_2 \left[|C_+|^2 v^{2i\mu}+|C_-|^2 v^{-2i\mu}\right] \prod_{i=1}^3
h_{\frac{l_i}{u}}(uv)h^*_{\frac{l_i}{u}}\left(\frac{u}{v}\right).
\end{equation*}
Similarly to the above discussion one can show that the product under this integral depends only on $v$. Hence, the calculations are essentially the same as we did for small external momenta. The only difference is in the argument of the logarithm and in the T-dependent prefactors. Hence,

\begin{equation}
n_2(k)  \approx \lambda^2\log \left(\frac{k\epsilon}{\mu} \right)\sigma^{int}(mT).
\end{equation}
Similarly for $\kappa_2(k)$ we use that:

\begin{equation*}
h^*_k(\eta_3)h^*_k(\eta_4)\approx C_+C_- \left[\left(\frac{\eta_3}{\eta_4} \right)^{i\mu}+\left(\frac{\eta_3}{\eta_4} \right)^{-i\mu} \right],
\end{equation*}
and, as a result,

\begin{equation}
\kappa_2(k) \propto -\lambda^2T^2k\int_{\frac{\mu}{k}}^\epsilon du\int^1 dv \iint \frac{dl_1 dl_2}{uv} C_+C_- \left(v^{2i\mu}+v^{-2i\mu} \right) \prod_{i=1}^3
h_{\frac{l_i}{u}}(uv)h^*_{\frac{l_i}{u}}\left(\frac{u}{v}\right).
\end{equation}
Here again as before, the product under the integral depends only on $v$. Hence, even for intermediate momenta we have logarithmically growing corrections both for $n$ and $\kappa$, $\kappa^*$.

At the same time, on general physical grounds it should be obvious that for high external momenta $k|\eta| \gg \mu$ we do not have any secular growth because the corresponding modes are not sensitive to the curvature of the space--time and behave as if they are in flat space. It is straightforward to show the latter fact explicitly (the situation is similar to that in flat space \cite{Akhmedov:2013vka}). Hence, below we concentrate on the small and intermediate parts of the external momenta, $k$, regions.

\section{Resummation of leading contributions from all loops in the expansion stage}

We see that even if $\lambda^2$ is small $\lambda^2 \log\left(\epsilon/T \right)$ and $\lambda^2 \log\left(k\epsilon/\mu\right)\approx \lambda^2 \log\left(k\epsilon/mT\right)$ can become large as $T/\epsilon \rightarrow + \infty$, hence quantum loop corrections are not suppressed in comparison with classical tree level contributions to the propagators. That evidently may strongly affect particle production. This means that we need to sum at least the leading corrections from all loops, and for that we have to use the system of Dyson-Schwinger (DS) equations \cite{Akhmedov:2013vka}. That is the mathematical part of the story.

Physically we have to keep in mind that the loop corrections to the Keldysh propagator of the previous section are found under assumption that $n$ and $\kappa$ do not change in time and keep their initial (vanishing in our case) values. But now we see that such an assumption is false if there are self--interactions and $T/\epsilon \to \infty$. One has to take the time evolution of $n$, $\kappa$ and $\kappa^*$ into account. That is what can be done with the use of the system of DS equations. The point is that Schwinger--Keldysh technique is causal and the system of equations under discussion actually provides two--point functions and vertexes as solutions of a Cauchy problem.

However, it is impossible to solve the system of the DS equations as it is. At least because it contains UV and subleading corrections on top of IR ones. But we can keep track only of the leading IR corrections from all loops. That drastically simplifies the system \cite{Akhmedov:2013vka}. In fact, once retarded, advanced propagators and vertexes receive only subleading corrections, which are suppressed by higher powers in $\lambda$, we can keep their tree--level values (perhaps UV renormalized, i.e. physical, as here we are taking care of only IR corrections). As a result, we have to deal only with the single equation for the Keldysh propagator.

Furthermore, because we are working here with the heavy fields, $mT \gg 1$, the modes are oscillatory functions for both small and large momenta, i.e. they oscillate even
when the system exits from the expansion stage. This results in the standard separation of scales between the time dependence of the modes $g_k(\eta)$ and that of particle density $n_k(\eta)$ and anomalous quantum average $\kappa_k(\eta)$ and $\kappa_k^*(\eta)$ \cite{LL}, \cite{Kamenev}. Namely $n$, $\kappa$ and $\kappa^*$ can be considered as very slow functions of time in comparison with $g_k(\eta)$. This fact also allows to simplify the DS equation for the Keldysh propagator. (Note that this is not the case for the light fields \cite{Akhmedov:2017ooy}.)

Thus, we assume that the quantum system under consideration had started its evolution at the vacuum state at the beginning of the de Sitter expansion stage, i.e. $n$ and $\kappa$ were remaining zero before $\eta=-T$. Hence, the tree-level or initial value of the Keldysh propagator $D^K_0$ is given by eq. (\ref{eq:19}) with $n_k = 0$, $\kappa_k = 0$ at $\eta=-T$. Our goal is to find the values of $D^K$ and of $n_k$, $\kappa_k$ at the exit from the expansion stage, i.e. at $\eta = - \epsilon$. Then, the relevant part of the system of DS equation takes the following form \cite{Akhmedov:2013vka}:

\begin{equation}\label{kineq1}
\begin{aligned}
{} & D^K(k|\eta_1,\eta_2) = D^K_0(k|\eta_1,\eta_2)+ \frac{\lambda^2}{6}\iint\frac{dq_1}{2\pi}\frac{dq_2}{2\pi}\iint \frac{d\eta_3 d\eta_4}{(\eta_3\eta_4)^2} \times \\ & \times \bigg[-\frac{1}{4}D^R_0(k|\eta_1,\eta_3)D^R_0(q_1|\eta_3,\eta_4)D^R_0(q_2|\eta_3,\eta_4)D^R_0(|k-q_1-q_2||\eta_3,\eta_4)D^K(k|\eta_4,\eta_2)+ \\& -\frac{1}{4}D^K(k|\eta_1,\eta_3)D^A_0(q_1|\eta_3,\eta_4)D^A_0(q_2|\eta_3,\eta_4)D^A_0(|k-q_1-q_2||\eta_3,\eta_4)D^A_0(k|\eta_4,\eta_2)+ \\ & -\frac{3}{4}D^R_0(k|\eta_1,\eta_3)D^K(q_1|\eta_3,\eta_4)D^R_0(q_2|\eta_3,\eta_4)D^R_0(|k-q_1-q_2||\eta_3,\eta_4)D^A_0(k|\eta_4,\eta_2)+ \\ & -\frac{3}{4}D^R_0(k|\eta_1,\eta_3)D^K(q_1|\eta_3,\eta_4)D^A_0(q_2|\eta_3,\eta_4)D^A_0(|k-q_1-q_2||\eta_3,\eta_4)D^A_0(k|\eta_4,\eta_2)+ \\ & +3D^R_0(k|\eta_1,\eta_3)D^K(q_1|\eta_3,\eta_4)D^K(q_2|\eta_3,\eta_4)D^R_0(|k-q_1-q_2||\eta_3,\eta_4)D^K(k|\eta_4,\eta_2)+ \\ & +3D^K(k|\eta_1,\eta_3)D^K(q_1|\eta_3,\eta_4)D^K(q_2|\eta_3,\eta_4)D^A_0(|k-q_1-q_2||\eta_3,\eta_4)D^A_0(k|\eta_4,\eta_2)+ \\ & +D^R_0(k|\eta_1,\eta_3)D^K(q_1|\eta_3,\eta_4)D^K(q_2|\eta_3,\eta_4)D^K(|k-q_1-q_2||\eta_3,\eta_4)D^A_0(k|\eta_4,\eta_2) \bigg],
\end{aligned}
\end{equation}
where $D^{R,A}_0$ are the initial (tree--level) values of the retarded and advanced propagators and $D^K$ is the exact Keldysh propagator whose form will be specified in a moment.

This equation is covariant under Bogoliubov rotations between different modes. For all the family of the modes which are related to each other via Bogolubov rotations \citep{Mottola:1984ar}, \citep{Allen:1985ux},  the following ansatz solves this equation \cite{Akhmedov:2013vka}:

\begin{equation}\label{ansatz}
D^K_k(\eta_1,\eta_2)=g_k(\eta_1)g^*_k(\eta_2)\left[\frac{1}{2} + n_k(\eta) \right] + g_k(\eta_1)g_k(\eta_2) \kappa_k(\eta) + c.c..
\end{equation}
Compare it to the eq. (\ref{eq:19}).

Above to simplify equations we have adopted a bit different notations. Namely we denote the exact $n$ and $\kappa$, $\kappa^*$ taken at a time $\eta$ as $n_k(\eta)$ and $\kappa_k(\eta)$. Note also that due to the approximate scale invariance, which is present in the expansion stage as a part of the de Sitter isometry (and which is violated only in the normalization factors of the modes), we can assume that $n_k(\eta) \approx n(k\eta) \equiv n_{k\eta}$ \cite{Akhmedov:2013vka}.

For all types of modes, which are related to each other via Bogolubov rotations, $n_k$ and $\kappa_k$ are comparable. The only exception are the out Jost modes $g_k \propto J_{i\mu}(k\eta)$, where $J_\nu(x)$ are the Bessel functions of the first kind. With the use of these Jost modes, the DS equation is solved by the previous ansatz with $\kappa_k = \kappa^* = 0$, if one considers only the leading terms in the limit under consideration \cite{Akhmedov:2013vka}.

What physical consequences do all these observations have? Before the start of the expansion, i.e. before $\eta = - T$, the theory under consideration is in its ground state, i.e. $n = 0 = \kappa = \kappa^*$ for the in--modes. During the expansion stage, between $\eta = - T$ and $\eta = - \epsilon$, $n$, $\kappa$ and $\kappa^*$ are generated and evolve in time. That describes the change of the state of the theory. The final state of the theory at the exit from the expansion stage is characterised by the values of $n$, $\kappa$ and $\kappa^*$ at $\eta = - \epsilon$. The latter values of $n$, $\kappa$ and $\kappa^*$ set up the initial quench for the reheating stage in the future Minkowski region, which will be discussed in the next section. At the same time, the generation of non--zero anomalous averages $\kappa$ and $\kappa^*$ just means that the initial ground state of the theory at past infinity is not anymore its ground state at future infinity.

Furthermore, the fact that the DS equation is covariant under Bogoliubov rotations between different modes and that for out--modes this equation has the solution with vanishing $\kappa$ and $\kappa^*$ is an evidence that the proper ground state of the theory is the out--vacuum \cite{Akhmedov:2013vka}.

All this means that to find a solution of the DS equation we have to perform the Bogoliubov rotation from in-- to the out--modes within the equation (\ref{kineq1}), plague there the ansatz (\ref{ansatz}) with $\kappa = \kappa^* = 0$, assume that $n$ is a slow function in comparison with $g_k$, take correspondingly $g_k$ to be the out--modes and, finally, pick up on the right hand side the leading contribution in the IR limit in question \cite{Akhmedov:2013vka}. The latter extraction is similar to that which have been done above for the second--loop correction. As the result of these manipulations we find the integro--differential equation:

\begin{equation}\label{kineq}
\begin{aligned}
{} & \frac{n_k(\eta)-n_k(T)}{\log(\eta)-\log(T)} \rightarrow \frac{dn_{k\eta}}{d\log(k\eta)} \approx -\frac{\lambda^2}{6mT}\iint\frac{dl_1}{2\pi}\frac{dl_2}{2\pi}\int\frac{dv}{v}\Big\{ 3{\rm Re} \left[v^{i\mu}\phi_{l_1}(v^{-1})\phi_{l_2}(v^{-1})\phi_{|l_1+l_2|}(v) \right] \times \\ & \times \left[(1+n_{k\eta})n_{l_1}n_{l_2}(1+n_{|l_1+l_2|})-n_{k\eta}(1+n_{l_1})(1+n_{l_2})n_{|l_1+l_2|} \right] + 3{\rm Re} \left[v^{i\mu}\phi_{l_1}(v^{-1})\phi_{l_2}(v)\phi_{|l_1+l_2|}(v) \right] \times \\ &\times \left[(1+n_{k\eta})n_{l_1}(1+n_{l_2})(1+n_{|l_1+l_2|}) - n_{k\eta}(1+n_{l_1})n_{l_2}n_{|l_1+l_2|} \right]+{\rm Re} \left[v^{i\mu}\phi_{l_1}(v^{-1})\phi_{l_2}(v^{-1})\phi_{|l_1+l_2|}(v^{-1}) \right] \times \\ & \times \left[(1+n_{k\eta})n_{l_1}n_{l_2}n_{|l_1+l_2|} - n_{k\eta}(1+n_{l_1})(1+n_{l_2})(1+n_{|l_1+l_2|}) \right]+{\rm Re} \left[v^{i\mu}\phi_{l_1}(v)\phi_{l_2}(v)\phi_{|l_1+l_2|}(v) \right] \times \\ &\times \left[(1+n_{k\eta})(1+n_{l_1})(1+n_{l_2})(1+n_{|l_1+l_2|}) - n_{k\eta} n_{l_1}n_{l_2}n_{|l_1+l_2|} \right] \Big\},
\end{aligned}
\end{equation}
where $l_i \equiv q_i\eta$ and we neglected $k$ in comparison with $q_i$, $i=1,2,3$, as that is the region where from the largest contributions are following \cite{Akhmedov:2013vka}. That is similar to the situation with the two--loop corrections.

For the low external momenta the out Jost modes behave as

\begin{equation}
\label{eq:45}
g_k(\eta) \approx F^s_k(k|\eta|)^{i\mu} = \frac{(kT)^{-i\mu}}{\sqrt{mT}}e^{-imT}(k|\eta|)^{i\mu}, \quad {\rm when} \quad mT \gg 1.
\end{equation}
This means that the function $\phi$ in eq. (\ref{kineq}) has the following form:

\[\phi_l(v) \propto
  \begin{cases}
  \frac{1}{mT}v^{i\mu}, \quad l\ll \mu & \text{}  \\
\frac{e^{-il\left(v-\frac{1}{v} \right)}}{l}, \quad l \gg \mu .  & \text{}
  \end{cases}
\]
The obtained eq. (\ref{kineq})  is valid for both low and intermediate external momenta.

First thing to observe is that the kinetic type equation (\ref{kineq}) does not possess Planckian distribution as its solution, because there is no energy conservation in time dependent backgrounds.

If we assume that the initial value of $n$ after the rotation to out Jost modes is very small\footnote{It should be much smaller than one, which is the case for the vanishing initial values of $n$, $\kappa$ and $\kappa^*$, if the mass $m$ is very big. Note that the initial values of $n$, $\kappa$ and $\kappa^*$ are taken to be vanishing for in--modes, but after the Bogoliubov rotation to the out--modes neither initial value of $n$ nor that of $\kappa$ and $\kappa^*$ does vanish any more. However, for large $mT$ they are much smaller than one. In such circumstances it was shown in \cite{Akhmedov:2017ooy} that $\kappa$ and $\kappa^*$ evolve to zero and $n$ solves the eq (\ref{kineq}).}, we can use the approximations adopted in \cite{Akhmedov:2013vka} to find the stationary solution of eq. (\ref{kineq}). If $n \ll 1$ the latter equation reduces to \cite{Akhmedov:2013vka}:

\begin{equation}
\frac{dn_{k\eta}}{d\log(k\eta)} \approx \Gamma_1 n_{k\eta} - \Gamma_2,
\end{equation}
where $\Gamma_1$ and $\Gamma_2$ are decay and production rates respectively, given by

\begin{equation*}
\Gamma_1=-\frac{\lambda^2}{6mT}\iint\frac{dl_1}{2\pi}\frac{dl_2}{2\pi}\int\frac{dv}{v} \, {\rm Re} \left\{v^{i\mu}\phi_{l_1}(v^{-1})\phi_{l_2}(v^{-1})\phi_{|l_1+l_2|}(v^{-1}) \right\},
\end{equation*}
\begin{equation}
\Gamma_2=-\frac{\lambda^2}{6mT}\iint\frac{dl_1}{2\pi}\frac{dl_2}{2\pi}\int\frac{dv}{v} \, {\rm Re} \left\{v^{i\mu}\phi_{l_1}(v)\phi_{l_2}(v)\phi_{|l_1+l_2|}(v) \right\}.
\end{equation}
Hence, for all $|k| < \mu/\epsilon$ we have flat stationary distribution:

\begin{equation}\label{distrib}
n(\eta = -\epsilon) \approx \frac{\Gamma_2}{\Gamma_1}.
\end{equation}
We can estimate the order of magnitude of this ratio if $\mu \approx mT \gg 1$ to verify the consistency of the result using the fact that the integrals are saturated at $l \sim \mu$, hence, neglecting the contributions from the remaining integration region \cite{Akhmedov:2013vka}. It is easy to see then that the so approximated integrand functions are just limits of products of Bessel functions, so we have the following estimate, in the limit $\frac{T}{\epsilon} \rightarrow +\infty$:

\begin{equation}
\label{eq:51}
\frac{\Gamma_2}{\Gamma_1} \approx e^{-3mT} \ll 1, \quad {\rm if} \quad mT \gg 1.
\end{equation}
This is the initial value for the dynamics in the future Minkowski region, which we discuss in the next section.

\section{Kinetic equation in the Minkowski out region}

Thus, at the exit from the expansion stage for all such momenta that $|k| < \mu/\epsilon$ we have the flat particle number density (\ref{distrib}), which is small, but non--zero. It is parametricaly small because we are dealing with large mass $mT \gg 1$. This distribution serves as an initial quench for the dynamics in the outer Minkowski region.  We expect an eventual thermalisation, because the latter region is flat and particle kinetics there is the same as in flat space. But we would like to see it from the first principles. That is what we will do in this section.

Above to calculate the tree--level expectation value of the stress--energy tensor we have glued the in--modes across $\eta = - \epsilon$ to find their behavior in the out Minkowski region. However, in the previous section we have observed that due to self--interactions the theory under consideration relaxes to the out--vacuum. Hence at the exit from the expansion stage the single quasi--particle states are represented by out--modes. As a result to proceed we have to glue the de Sitter out--modes rather than in--modes across $\eta = - \epsilon$. Namely, to find the form of the modes in the out Minkowskian region we have to glue the solution in the out flat region with the out Jost modes from the de Sitter stage.

For small momenta, $k|\eta| \ll \mu$, the gluing conditions give the following system of equations

\[
  \begin{cases}
 \sqrt{\epsilon}F^s_k\left(k\epsilon \right)^{i\mu}=\frac{1}{\sqrt{\omega_{out}}} \left(C^s_k e^{i\omega_{out}\epsilon}+D^s_k e^{-i\omega_{out}\epsilon} \right) & \text{}  \\
 i \mu k\sqrt{\epsilon} F^s_k \left( k\epsilon \right)^{i\mu-1}+\frac{1}{2\sqrt{\epsilon}}F^s_k \left(k\epsilon \right)^{i\mu}=i\sqrt{\omega_{out}}\left(C^s_k e^{i\omega_{out}\epsilon}-D^s_k e^{-i\omega_{out}\epsilon} \right).  & \text{}
  \end{cases}
\]\\
Solving this system we find the coefficients

\begin{equation*}
C^s_k=\frac{e^{-i\omega_{out}\epsilon}}{2}F^s_k \left(k\epsilon \right)^{i\mu}\left[\sqrt{\omega_{out}\epsilon}+\frac{\mu}{\sqrt{\omega_{out}\epsilon}}-\frac{i}{2\sqrt{\omega_{out}\epsilon}} \right],
\end{equation*}
\begin{equation}
D^s_k= \frac{e^{i\omega_{out}\epsilon}}{2}F^s_k \left(k\epsilon \right)^{i\mu} \left[\sqrt{\omega_{out}\epsilon}-\frac{\mu}{\sqrt{\omega_{out}\epsilon}}+\frac{i}{2\sqrt{\omega_{out}\epsilon}} \right],
\end{equation}
and with the usual approximations $\omega_{in}\approx m$, $\omega_{out}\approx \frac{mT}{\epsilon}$ and $\mu \approx mT$ we have the following estimates

\begin{equation}
C^s_k \approx e^{-2imT}\left(\frac{T}{\epsilon} \right)^{-imT}, \quad {\rm and} \quad D^s_k \approx 0
\end{equation}
so the modes in the out region behave as single waves


\begin{equation}
g_k(t) \propto \sqrt{\frac{\epsilon}{mT}} e^{-2imT}\left(\frac{T}{\epsilon} \right)^{-imT} e^{-imt},
\end{equation}
where we also used the explicit form of the coefficient of the Jost function, $F^s_k$, calculated in the previous section in eq. (\ref{eq:45}).

Let us now apply the same gluing conditions for the modes of intermediate momenta ($k\epsilon \ll \mu \ll kT$)

\begin{equation*}
C^s_k=\frac{e^{-i\omega_{out}\epsilon}}{2}F^{int}_k \left(k\epsilon \right)^{i\mu}\left[\sqrt{\omega_{out}\epsilon}+\frac{\mu}{\sqrt{\omega_{out}\epsilon}}-\frac{i}{2\sqrt{\omega_{out}\epsilon}} \right],
\end{equation*}
\begin{equation}
D^s_k= \frac{e^{i\omega_{out}\epsilon}}{2}F^{int}_k \left(k\epsilon \right)^{i\mu} \left[\sqrt{\omega_{out}\epsilon}-\frac{\mu}{\sqrt{\omega_{out}\epsilon}}+\frac{i}{2\sqrt{\omega_{out}\epsilon}} \right].
\end{equation}
As a result


\begin{equation}
g_k(t) \propto \frac{1}{\sqrt{\omega_{out}(k)}} e^{-2ikT}\left(k\epsilon \right)^{imT} e^{-imt}.
\end{equation}
We have seen that the particle density created during the de Sitter expansion stage is not Planckian (\ref{eq:51}).
Hence, our purpose now is to find a kinetic equation for $n_k(t)$ in the out Minkowski region using as initial value for $n$ the one from (\ref{eq:51}). The point is that if the initial density is not Plankian, i.e. is not stationary for flat space kinetic equation, it should evolve in time towards the stationary Plankian distribution. To see that, essentially we have to perform similar calculations to those we already did to find the kinetic equation during the de Sitter expansion stage, but using this time the modes we found in this section above. Note that in general anomalous average $\kappa_k$ also grows along with $n_k$, but because the modes under consideration represent single waves, it is not hard to see that $\kappa_k \approx 0$ \cite{Akhmedov:2013vka}.

Let us write the modes in this out region in the following way:

\begin{equation}\label{gk}
g_k(t) \approx \frac{1}{\sqrt{\omega(k)}}Q_{T,\epsilon} e^{-i\omega(k)t}.
\end{equation}
As we will see this form will be good for both regions of momenta, considering $|Q_{T,\epsilon}|^2 = 1$ in both small and intermediate momenta case.
From now on, for the sake of notational simplicity we will write just $\omega$ having in mind we are talking about $\omega_{out}$.

We can follow the same approach as was already used above, just using (\ref{gk}) instead of Bessel functions and considering that the loop corrections to $n_k$ at leading order grow linearly with time, to derive the following equation for the time evolution of $n_k$:

\begin{equation}
\begin{aligned}
{} & \frac{dn_k(t)}{dt}  \propto \frac{\lambda^2}{\omega(k)}\iint \frac{dq_1 dq_2}{\omega(q_1)\omega(q_2)\omega(k-q_1-q_2)} \times  \\ & \times \bigg\{3\int_{-\epsilon}^t \cos\left[\left(-\omega(k)-\omega(q_1)+\omega(q_2)+\omega(k+q_1-q_2) \right)(t+\epsilon) \right] \times \\ & \times \left[(1+n_k)(1+n_{q_1})n_{q_2}n_{|k+q_1-q_2|}-n_k n_{q_1}(1+n_{q_2})(1+n_{|k+q_1-q_2|}) \right] + \\ & +3\int_{-\epsilon}^t \cos\left[\left(\omega(k)-\omega(q_1)+\omega(q_2)+\omega(-k+q_1-q_2) \right)(t+\epsilon) \right] \times \\ & \times \left[(1+n_k)n_{q_1}(1+n_{q_2})(1+n_{|-k+q_1-q_2|})-n_k(1+n_{q_1})n_{q_2}n_{|-k+q_1-q_2|} \right] + \\ & +\int_{-\epsilon}^t \cos\left[\left(-\omega(k)+\omega(q_1)+\omega(q_2)+\omega(k-q_1-q_2) \right)(t+\epsilon) \right] \times \\ & \times \left[(1+n_k)n_{q_1}n_{q_2}n_{|k-q_1-q_2|}-n_k(1+n_{q_1})(1+n_{q_2})(1+n_{|k-q_1-q_2|}) \right] + \\ & +\int_{-\epsilon}^t \cos\left[\left(\omega(k)+\omega(q_1)+\omega(q_2)+\omega(k+q_1+q_2) \right)(t+\epsilon) \right] \times \\ & \times \left[(1+n_k)(1+n_{q_1})(1+n_{q_2})(1+n_{|k+q_1+q_2|})-n_k n_{q_1}n_{q_2}n_{|k-q_1-q_2|} \right]
\bigg\}.
\end{aligned}
\end{equation}
Computing the cosine time integrals we obtain terms of the form $\frac{\sin\left[\Delta\omega (t+\epsilon) \right]}{\Delta\omega}$ and in the limit of $t \rightarrow +\infty$ these are reduced to $\delta$-functions ensuring energy conservation, as it should be the case in flat space.
The only allowed process is then the scattering one between scalar particles, so the collision integral of the kinetic equation just contain the term with $(1+n_k)n_{q_1}n_{q_2}(1+n_{|k-q_1-q_2|})-n_k(1+n_{q_1})(1+n_{q_2})n_{|k-q_1-q_2|}$.
Given that our initial density is small, $n_k \ll 1$, we can approximate the expression as $n_{q_1}n_{q_2}-n_kn_{|k-q_1-q_2|}$.

After these manipulations and approximations we are left with the standard Boltzmann equation

\begin{eqnarray}
\frac{dn_k(t)}{dt} \propto \frac{\lambda^2}{\omega(k)}\iint \frac{dq_1 dq_2}{\omega(q_1)\omega(q_2)\omega(k-q_1-q_2)}\times \nonumber \\ \times \left[n_{q_1}n_{q_2}-n_kn_{|k-q_1-q_2|} \right]\delta\left(\omega(k)-\omega(q_1)-\omega(q_2)+\omega(k-q_1-q_2) \right)
\end{eqnarray}
and we can immediately see that the stationary solution is given by the equilibrium Boltzmann distribution $n_k \propto e^{-\frac{\omega(k)}{\tau}}$ for a constant $\tau$, resulting in thermalization. The value of this constant can be estimated from the total energy that was stored by the particles of the density (\ref{eq:51}) at the moment of exit from the expansion stage. I.e. we should have an equality $\int_{-\mu/\epsilon}^{\mu/\epsilon} n \, \omega_{out}(k) dk \approx \tau$, hence,

 \[
 \boxed{\tau \sim \frac{mT}{\epsilon} \, e^{-3mT}}.
 \]

\section{Conclusions}

In the section 3 we have calculated the stress--energy expectation value due to the zero point fluctuations. That is due to $1/2$ term in eq. (\ref{eq:19}), i.e. when $n$, $\kappa$ and $\kappa^*$ are vanishing. In the section 4 we have shown that quantum loop corrections to $n$, $\kappa$ and $\kappa^*$ are not negligible, if the expansion stage is long enough. This signals the breakdown of the perturbation theory or semi--classical approximation and calls for the resummation of at least the leading corrections from all loops. That is done in the section 5. The generation of non--zero $n$, $\kappa$ and $\kappa^*$ affects the picture observed in the section 3. At the exit from the expansion stage we have a non--zero physical particle number density even for very massive fields. The latter one is thermalized in the future Minkowski region, as is shown in the section 6.

In this note we have considered very massive fields to have an analytical headway. For the massive fields the backreaction on the background gravitational field is negligible. That is because the comoving density (\ref{distrib}) remains finite, which means that the physical one gets diluted as the spatial volume increases during the expansion. As a result, the expectation value of the stress--energy tensor due to created particles has a negligible
effect on the background geometry.

However, in \cite{Akhmedov:2013vka} it is shown that even for massive fields the kinetic equation (\ref{kineq}) has explosive solution with a dense enough initial state. The problem with such a situation is that for the latter value of the initial particle number density the matter energy density is comparable or even grater than the cosmological constant and, hence, the discovery of the explosive solution is not consistent.

However, for the light particles the situation is quite different, as is shown in \cite{Akhmedov:2017ooy}. In the latter case the explosive solution of the corresponding substitution of the kinetic equation appears even when the initial matter energy density is much smaller than the cosmological constant. Moreover, light particles are much more interesting from the phenomenological point of view. Hence, we plan to discuss light particles elsewhere. But from the computational point of view the consideration of the light particles is much more complicated than that of the massive particles. Apart from all it demands a consideration of the coupled system of the gravitational field and of the matter, because in such a case the backreaction on the gravitational background can be non--negligible \cite{Akhmedov:2017ooy}.

We would like to acknowledge discussions with S.Alexeev and L.Astrakhantsev.
The work of ETA was done under the partial support of the RFBR grant 16-02-01021 and of the state grant Goszadanie 3.9904.2017/8.9.


\begin{thebibliography}{99}

\bibitem{Kofman:1994rk}
  L.~Kofman, A.~D.~Linde and A.~A.~Starobinsky,
  Phys.\ Rev.\ Lett.\  {\bf 73}, 3195 (1994)
  doi:10.1103/PhysRevLett.73.3195
  [hep-th/9405187].

\bibitem{Starobinsky:1980te}
  A.~A.~Starobinsky,
  Phys.\ Lett.\  {\bf 91B}, 99 (1980).
  doi:10.1016/0370-2693(80)90670-X

\bibitem{Guth:1980zm}
  A.~H.~Guth,
  Phys.\ Rev.\ D {\bf 23}, 347 (1981).
  doi:10.1103/PhysRevD.23.347

\bibitem{Linde:1981mu}
  A.~D.~Linde,
  Phys.\ Lett.\  {\bf 108B}, 389 (1982).
  doi:10.1016/0370-2693(82)91219-9

\bibitem{Albrecht:1982wi}
  A.~Albrecht and P.~J.~Steinhardt,
  Phys.\ Rev.\ Lett.\  {\bf 48}, 1220 (1982).
  doi:10.1103/PhysRevLett.48.1220

\bibitem{MarolfMorrison} 
  D.~Marolf, I.~A.~Morrison and M.~Srednicki,
  ``Perturbative S-matrix for massive scalar fields in global de Sitter space,''
  arXiv:1209.6039 [hep-th].

\bibitem{Marolf:2010zp}
  D.~Marolf and I.~A.~Morrison,
  Phys.\ Rev.\ D {\bf 82}, 105032 (2010)
  [arXiv:1006.0035 [gr-qc]].

\bibitem{Marolf:2011sh}
  D.~Marolf and I.~A.~Morrison,
  Gen.\ Rel.\ Grav.\  {\bf 43}, 3497 (2011)
  [arXiv:1104.4343 [gr-qc]].

\bibitem{Higuchi:2010xt}
  A.~Higuchi, D.~Marolf, I.~A.~Morrison,
  Phys.\ Rev.\  {\bf D83}, 084029 (2011).
  [arXiv:1012.3415 [gr-qc]].

\bibitem{Marolf:2010nz}
  D.~Marolf and I.~A.~Morrison,
  Phys.\ Rev.\ D {\bf 84}, 044040 (2011)
  [arXiv:1010.5327 [gr-qc]].

\bibitem{Hollands:2010pr}
  S.~Hollands,
  Commun.\ Math.\ Phys.\  {\bf 319}, 1 (2013)
  [arXiv:1010.5367 [gr-qc]].

\bibitem{Hollands:2011we}
  S.~Hollands,
  Annales Henri Poincare {\bf 13}, 1039 (2012)
  [arXiv:1105.1996 [gr-qc]].

\bibitem{Dolgov:1994cq}
  A.~D.~Dolgov, M.~B.~Einhorn and V.~I.~Zakharov,
  Phys.\ Rev.\  D {\bf 52}, 717 (1995)
  [arXiv:gr-qc/9403056].

\bibitem{Dolgov:1994ra}
  A.~D.~Dolgov, M.~B.~Einhorn and V.~I.~Zakharov,
  Acta Phys.\ Polon.\  B {\bf 26}, 65 (1995)
  [arXiv:gr-qc/9405026].

\bibitem{Burgess:2010dd}
  C.~P.~Burgess, R.~Holman, L.~Leblond and S.~Shandera,
  JCAP {\bf 1010}, 017 (2010)
  [arXiv:1005.3551 [hep-th]].

\bibitem{Burgess:2009bs}
  C.~P.~Burgess, L.~Leblond, R.~Holman and S.~Shandera,
  JCAP {\bf 1003}, 033 (2010)
  [arXiv:0912.1608 [hep-th]].

\bibitem{Giddings:2010ui}
  S.~B.~Giddings, M.~S.~Sloth,
  JCAP {\bf 1007}, 015 (2010).
  [arXiv:1005.3287 [hep-th]].

\bibitem{Giddings:2010nc}
  S.~B.~Giddings, M.~S.~Sloth,
  JCAP {\bf 1101}, 023 (2011).
  [arXiv:1005.1056 [hep-th]].

\bibitem{Riotto:2008mv}
  A.~Riotto, M.~S.~Sloth,
  JCAP {\bf 0804}, 030 (2008).
  [arXiv:0801.1845 [hep-ph]].

\bibitem{Onemli:2002hr}
  V.~K.~Onemli, R.~P.~Woodard,
  Class.\ Quant.\ Grav.\  {\bf 19}, 4607 (2002).
  [gr-qc/0204065].

\bibitem{Onemli:2004mb}
  V.~K.~Onemli, R.~P.~Woodard,
  Phys.\ Rev.\  {\bf D70}, 107301 (2004).
  [gr-qc/0406098].

\bibitem{Kahya:2006hc}
  E.~O.~Kahya, V.~K.~Onemli,
  Phys.\ Rev.\  {\bf D76}, 043512 (2007).
  [gr-qc/0612026].

\bibitem{Kahya:2009sz}
  E.~O.~Kahya, V.~K.~Onemli, R.~P.~Woodard,
  Phys.\ Rev.\  {\bf D81}, 023508 (2010).
  [arXiv:0904.4811 [gr-qc]].

\bibitem{Tsamis:2005hd}
  N.~C.~Tsamis and R.~P.~Woodard,
  Nucl.\ Phys.\ B {\bf 724}, 295 (2005)
  doi:10.1016/j.nuclphysb.2005.06.031
  [gr-qc/0505115].

\bibitem{Xue:2011hm}
  W.~Xue, K.~Dasgupta, R.~Brandenberger,
  Phys.\ Rev.\  {\bf D83}, 083520 (2011).
  [arXiv:1103.0285 [hep-th]].

\bibitem{Woodard:2014jba}
  R.~P.~Woodard,
  Int.\ J.\ Mod.\ Phys.\ D {\bf 23}, no. 09, 1430020 (2014)
  doi:10.1142/S0218271814300201
  [arXiv:1407.4748 [gr-qc]].

\bibitem{Garbrecht:2013coa}
  B.~Garbrecht, G.~Rigopoulos and Y.~Zhu,
  Phys.\ Rev.\ D {\bf 89}, 063506 (2014)
  doi:10.1103/PhysRevD.89.063506
  [arXiv:1310.0367 [hep-th]].

\bibitem{Boyanovsky:2015xoa}
  D.~Boyanovsky,
  New J.\ Phys.\  {\bf 17}, no. 6, 063017 (2015)
  doi:10.1088/1367-2630/17/6/063017
  [arXiv:1503.00156 [hep-ph]].

\bibitem{Boyanovsky:2015tba}
  D.~Boyanovsky,
  Phys.\ Rev.\ D {\bf 92}, no. 2, 023527 (2015)
  doi:10.1103/PhysRevD.92.023527
  [arXiv:1506.07395 [astro-ph.CO]].

\bibitem{Lello:2013qza}
  L.~Lello, D.~Boyanovsky and R.~Holman,
  JHEP {\bf 1404}, 055 (2014)
  doi:10.1007/JHEP04(2014)055
  [arXiv:1305.2441 [astro-ph.CO]].

\bibitem{Benedetti:2014gja}
  D.~Benedetti,
  J.\ Stat.\ Mech.\  {\bf 1501}, P01002 (2015)
  doi:10.1088/1742-5468/2015/01/P01002
  [arXiv:1403.6712 [cond-mat.stat-mech]].

\bibitem{Boyanovsky:2014uya}
  D.~Boyanovsky and L.~Lello,
  New J.\ Phys.\  {\bf 16}, 063050 (2014)
  doi:10.1088/1367-2630/16/6/063050
  [arXiv:1403.6366 [hep-ph]].

\bibitem{Boyanovsky:2015jen}
  D.~Boyanovsky,
  Phys.\ Rev.\ D {\bf 93}, 043501 (2016)
  doi:10.1103/PhysRevD.93.043501
  [arXiv:1511.06649 [astro-ph.CO]].

\bibitem{Nacir:2016fzi}
  D.~Lopez Nacir, F.~D.~Mazzitelli and L.~G.~Trombetta,
  JHEP {\bf 1609}, 117 (2016)
  doi:10.1007/JHEP09(2016)117
  [arXiv:1606.03481 [hep-th]].

\bibitem{Onemli:2015tca}
  V.~K.~Onemli,
  arXiv:1510.02272 [gr-qc].

\bibitem{LopezNacir:2016gfj}
  D.~Lopez Nacir, F.~D.~Mazzitelli and L.~G.~Trombetta,
  EPJ Web Conf.\  {\bf 125}, 05019 (2016)
  doi:10.1051/epjconf/201612505019
  [arXiv:1610.09943 [hep-th]].

\bibitem{Boyanovsky:1993xf}
  D.~Boyanovsky, H.~J.~de Vega and R.~Holman,
  Phys.\ Rev.\ D {\bf 49}, 2769 (1994)
  doi:10.1103/PhysRevD.49.2769
  [hep-ph/9310319].

\bibitem{deVega:1997yw}
  H.~J.~de Vega and J.~F.~J.~Salgado,
  Phys.\ Rev.\ D {\bf 56}, 6524 (1997)
  doi:10.1103/PhysRevD.56.6524
  [hep-th/9701104].

\bibitem{Boyanovsky:1997cr}
  D.~Boyanovsky, D.~Cormier, H.~J.~de Vega, R.~Holman, A.~Singh and M.~Srednicki,
  Phys.\ Rev.\ D {\bf 56}, 1939 (1997)
  doi:10.1103/PhysRevD.56.1939
  [hep-ph/9703327].

\bibitem{Onemli:2015pma}
  V.~K.~Onemli,
  Phys.\ Rev.\ D {\bf 91}, 103537 (2015)
  doi:10.1103/PhysRevD.91.103537
  [arXiv:1501.05852 [gr-qc]].

\bibitem{Onemli:2013gya}
  V.~K.~Onemli,
  Phys.\ Rev.\ D {\bf 89}, 083537 (2014)
  doi:10.1103/PhysRevD.89.083537
  [arXiv:1312.6409 [astro-ph.CO]].

\bibitem{Gautier:2015pca}
  F.~Gautier and J.~Serreau,
  Phys.\ Rev.\ D {\bf 92}, no. 10, 105035 (2015)
  doi:10.1103/PhysRevD.92.105035
  [arXiv:1509.05546 [hep-th]].

\bibitem{Guilleux:2015pma}
  M.~Guilleux and J.~Serreau,
  Phys.\ Rev.\ D {\bf 92}, no. 8, 084010 (2015)
  doi:10.1103/PhysRevD.92.084010
  [arXiv:1506.06183 [hep-th]].

\bibitem{Serreau:2013eoa}
  J.~Serreau,
  Phys.\ Lett.\ B {\bf 730}, 271 (2014)
  doi:10.1016/j.physletb.2014.01.058
  [arXiv:1306.3846 [hep-th]].

\bibitem{Gautier:2013aoa}
  F.~Gautier and J.~Serreau,
  Phys.\ Lett.\ B {\bf 727}, 541 (2013)
  doi:10.1016/j.physletb.2013.10.072
  [arXiv:1305.5705 [hep-th]].

\bibitem{Guilleux:2016oqv}
  M.~Guilleux and J.~Serreau,
  arXiv:1611.08106 [gr-qc].

\bibitem{Serreau:2013psa}
  J.~Serreau and R.~Parentani,
  Phys.\ Rev.\ D {\bf 87}, 085012 (2013)
  doi:10.1103/PhysRevD.87.085012
  [arXiv:1302.3262 [hep-th]].

\bibitem{PolyakovKrotov}
  D.~Krotov, A.~M.~Polyakov,
  Nucl.\ Phys.\  {\bf B849}, 410-432 (2011).
  [arXiv:1012.2107 [hep-th]].

\bibitem{Polyakov:2012uc}
  A.~M.~Polyakov,
  ``Infrared instability of the de Sitter space,''
  arXiv:1209.4135 [hep-th].

\bibitem{Polyakov:2007mm}
  A.~M.~Polyakov,
  Nucl.\ Phys.\  B {\bf 797}, 199 (2008)
  [arXiv:0709.2899 [hep-th]].

\bibitem{Polyakov:2009nq}
  A.~M.~Polyakov,
  Nucl.\ Phys.\  B {\bf 834}, 316 (2010)
  [arXiv:0912.5503 [hep-th]].

\bibitem{AkhmedovKEQ}
  E.~T.~Akhmedov,
  JHEP {\bf 1201}, 066 (2012)
  [arXiv:1110.2257 [hep-th]].

\bibitem{AkhmedovBurda}
  E.~T.~Akhmedov and P.~.Burda,
  Phys.\ Rev.\ D {\bf 86}, 044031 (2012)
  [arXiv:1202.1202 [hep-th]].

\bibitem{AkhmedovGlobal}
  E.~T.~Akhmedov,
  Phys.\ Rev.\ D {\bf 87}, 044049 (2013)
  [arXiv:1209.4448 [hep-th]].

\bibitem{AkhmedovPopovSlepukhin}
  E.~T.~Akhmedov, F.~K.~Popov and V.~M.~Slepukhin,
  Phys.\ Rev.\ D {\bf 88}, 024021 (2013)
  [arXiv:1303.1068 [hep-th]].

\bibitem{Akhmedov:2017ooy}
  E.~T.~Akhmedov, U.~Moschella, K.~E.~Pavlenko and F.~K.~Popov,
  arXiv:1701.07226 [hep-th].


\bibitem{Akhmedov:2013vka}
  E.~T.~Akhmedov,
  International Journal of Modern Physics D, Vol.\  23, No.\  {\bf 1}, 1430001 (2014)
  [arXiv:1309.2557 [hep-th]].

\bibitem{LL}
L.~D.~Landau and E.~M.~Lifshitz, Vol. 10 (Pergamon Press, Oxford, 1975).

\bibitem{Kamenev} A.Kamenev, ``Many-body theory of non-equilibrium systems'',  arXiv:cond-mat/0412296;
Bibliographic Code:	2004cond.mat.12296K.


\bibitem{Bunch:1978yq}
  T.~S.~Bunch and P.~C.~W.~Davies,
  Proc.\ Roy.\ Soc.\ Lond.\  A {\bf 360}, 117 (1978).


\bibitem{Akhmedov:2014hfa}
  E.~T.~Akhmedov, N.~Astrakhantsev and F.~K.~Popov,
  JHEP {\bf 1409}, 071 (2014)
  doi:10.1007/JHEP09(2014)071
  [arXiv:1405.5285 [hep-th]].

\bibitem{Akhmedov:2014doa}
  E.~T.~Akhmedov and F.~K.~Popov,
  JHEP {\bf 1509}, 085 (2015)
  doi:10.1007/JHEP09(2015)085
  [arXiv:1412.1554 [hep-th]].

\bibitem{Akhmedov:2015xwa}
  E.~T.~Akhmedov, H.~Godazgar and F.~K.~Popov,
  Phys.\ Rev.\ D {\bf 93}, no. 2, 024029 (2016)
  doi:10.1103/PhysRevD.93.024029
  [arXiv:1508.07500 [hep-th]].

\bibitem{Akhmedov:2017hbj}
  E.~T.~Akhmedov and S.~O.~Alexeev,
  Phys.\ Rev.\ D {\bf 96}, no. 6, 065001 (2017)
  doi:10.1103/PhysRevD.96.065001
  [arXiv:1707.02242 [hep-th]].

\bibitem{Alexeev:2017ath}
  S.~Alexeev,
  arXiv:1707.02838 [hep-th].

  \bibitem{BirDav}
  N.~D.~Birrell and P.~C.~W.~Davies, "Quantum Fields in Curved Space" (Cambridge University Press, 1984),
  doi:10.1017/CBO9780511622632


\bibitem{vanderMeulen:2007ah}
  M.~van der Meulen, J.~Smit,
  JCAP {\bf 0711}, 023 (2007).
  [arXiv:0707.0842 [hep-th]].

\bibitem{Allen:1985ux}
  B.~Allen,
  ``Vacuum States In De Sitter Space,''
  Phys.\ Rev.\  D {\bf 32}, 3136 (1985).

\bibitem{Mottola:1984ar}
  E.~Mottola,
  Phys.\ Rev.\  D {\bf 31}, 754 (1985).



\end{thebibliography}

\end{document}